\documentclass[iop]{emulateapj}

\newcommand{\secpoint}{${.}\!\!^{\prime \prime}\!$}
\newcommand{\minpoint}{${.}\!\!^{\prime}$}
\newcommand{\etal}{et~al.\ }
\newcommand{\ie}{i.e.,\ }
\newcommand{\eg}{e.g.\ }
\newcommand{\deli}{$\Delta i'$\ }

\newcommand{\gio}{$(g'-i')_o$\ }
\newcommand{\go}{$g'_{o}$\ }
\newcommand{\io}{$i'_{o}$\ }
\newcommand{\bgc}{$\Sigma_{bGC}$\ }
\newcommand{\gc}{$\Sigma_{GC}$\ }

\newcommand{\rgc}{$\Sigma_{rGC}$\ }

\newcommand{\bgct}{$\Sigma_{bGC,tot}$\ }

\slugcomment{GCs in the Virgo Cluster}

\shorttitle{GCs in the Virgo Cluster}

\shortauthors{Durrell et al.}

\begin{document}

\title{The Next Generation Virgo Cluster Survey. VIII. The Spatial Distribution of Globular Clusters in the Virgo Cluster}

\author{Patrick R. Durrell$^{1}$, Patrick C{\^o}t{\'e}$^{2}$, Eric W. Peng$^{3,4}$, John P. Blakeslee$^{2}$, Laura Ferrarese$^{2},$ J.Christopher Mihos$^{5}$, Thomas H. Puzia$^{6}$, Ariane Lan\c{c}on$^{7}$, Chengze Liu$^{8,9}$, Hongxin Zhang$^{3,4}$, Jean-Charles Cuillandre$^{10}$, Alan McConnachie$^{2}$,  Andr{\'e}s Jord{\'a}n$^{6}$, Katharine Accetta$^{1}$, Samuel Boissier$^{11}$, Alessandro Boselli$^{11}$, St{\'e}phane Courteau$^{12}$, Pierre-Alain Duc$^{13}$, Eric Emsellem$^{14,15}$, Stephen Gwyn$^{2}$, Simona Mei$^{16,17}$, and James E. Taylor$^{18}$}

\affil{$^1$Department of Physics \& Astronomy, Youngstown State University, Youngstown, OH USA 44555}
\affil{$^2$Herzberg Astronomy and Astrophysics, National Research Council,  5071 W. Saanich Road, Victoria, BC, V9E 2E7, Canada}
\affil{$^3$Department of Astronomy, Peking University, Beijing 100871, China}
\affil{$^4$Kavli Institute for Astronomy and Astrophysics, Peking University, Beijing 100871, China }
\affil{$^5$Department of Astronomy, Case Western Reserve University, Cleveland, OH, 44106, USA}
\affil{$^6$Institute of Astrophysics, Pontificia Universidad Catolica, Av. Vicu–a Mackenna 4860, Macul 7820436, Santiago, Chile}
\affil{$^7$Observatoire astronomique de Strasbourg, Universit{\'e} de Strasbourg, CNRS, UMR 7550, 11 rue de lÕUniversit{\'e}, F-67000 Strasbourg, France}
\affil{$^8$Center for Astronomy and Astrophysics, Department of Physics and Astronomy, Shanghai Jiao Tong University, 800 Dongchuan Road, Shanghai 200240, China}
\affil{$^{9}$Shanghai Key Lab for Particle Physics and Cosmology, Shanghai Jiao Tong University, Shanghai, 200240, China}
\affil{$^{10}$Canada-France-Hawaii Telescope Corporation, Kamuela HI 96743, USA}
\affil{$^{11}$Aix Marseille Universit{\'e}, CNRS, LAM (Laboratoire d'Astrophysique de Marseille) UMR 7326, F-13388, Marseille, France}
\affil{$^{12}$Department of Physics, Engineering Physics and Astronomy, QueenÕs University, Kingston, ON, K7L 3N6, Canada}
\affil{$^{13}$AIM Paris Saclay, CNRS/INSU, CEA/Irfu, Universit«e Paris Diderot, Orme des Merisiers, F-91191 Gif sur Yvette cedex, France}
\affil{$^{14}$Universit{\'e} de Lyon 1, CRAL, Observatoire de Lyon, 9 av. Charles Andr{\'e}, F-69230 Saint-Genis Laval; CNRS, UMR 5574; ENS de Lyon, France}
\affil{$^{15}$European Southern Observatory, Karl-Schwarzschild-Str. 2, D-85748 Garching, Germany}
\affil{$^{16}$GEPI, Observatoire de Paris, Section de Meudon, 5 Place J. Janssen, 92190 Meudon Cedex, France}
\affil{$^{17}$Universit\'{e} Paris Denis Diderot, 75205 Paris Cedex 13, France}
\affil{$^{18}$University of Waterloo, Waterloo, ON, N2L 3G1 Canada}

\begin{abstract}

We report on a large-scale study of the distribution of globular clusters (GCs) throughout the Virgo cluster, based on photometry from the Next Generation Virgo Cluster Survey (NGVS), a large imaging survey covering Virgo's primary subclusters (Virgo A=M87 and Virgo B=M49) out to their virial radii.  Using the $g'_{o}$, \gio color-magnitude diagram of unresolved and marginally resolved sources within the NGVS, we have constructed 2-D maps of the (irregular) GC distribution over 100 square degrees to a depth of \go=24.  We present the clearest evidence to date showing the difference in concentration between red and blue GCs over the full extent of the cluster, where the red (more metal-rich) GCs are largely located around the massive early-type galaxies in Virgo, while the blue (metal-poor) GCs have a much more extended spatial distribution, with significant populations still present beyond 83' ($\sim 215$ kpc) along the major axes of both M49 and M87.  A comparison of our GC maps to the diffuse light in the outermost regions of M49 and M87 show remarkable agreement in the shape, ellipticity, and boxiness of both luminous systems.  We also find evidence for spatial enhancements of GCs surrounding M87 that may be indicative of recent interactions or an ongoing merger history.    We compare the GC map to that of the locations of Virgo galaxies and the X-ray intracluster gas, and find generally good agreement between these various baryonic structures.  We calculate the Virgo cluster contains a total population of $N_{GC}=67,300\pm 14,400$, of which 35\% are located in M87 and M49 alone.  For the first time, we compute a cluster-wide specific frequency $S_{N,CL}=2.8\pm 0.7$, after correcting for Virgo's diffuse light.    We also find a GC-to-baryonic mass fraction $\epsilon_b=5.7\pm1.1 \times 10^{-4}$ and a GC-to-total cluster mass formation efficiency $\epsilon_t=2.9\pm0.5 \times 10^{-5}$, the latter values slightly lower than, but consistent with, those derived for individual galactic halos.   Taken as a whole, our results show that the production of the complex structures in the unrelaxed Virgo cluster core (including the production of the diffuse intracluster light) is an ongoing and continuing process.  

\end{abstract}


\section{Introduction}

\setcounter{footnote}{0}

A priority in extragalactic astrophysics is to understand the
hierarchical assembly of galaxy clusters, which are expected (in the
$\Lambda$CDM paradigm) to have built up over time, with the larger
galaxies (especially the brightest cluster galaxies, BCGs) having
grown through mergers and accretion of many smaller galaxies\footnote{The timescales over which this happens, however, is
still under debate \citep[\eg][]{col09}}\citep[\eg][]{dub98,spring05,naab09,rs09,oser10,lap13}.
Thus an understanding of the mass assembly of galaxy clusters is
heavily dependent on the large number of galactic interactions that
occur over the entire history of the cluster.  Detailed studies of the
galaxies, intracluster gas, and discrete stellar
populations (\ie stars, nebulae, star clusters) within the cluster
galaxies have all provided insights into the basic framework of
hierarchical evolution.

The buildup of the extended stellar halo around BCGs
(and perhaps other massive cluster members) from interaction with 
merged/stripped galaxies also results in the 
formation of a much more extended envelope of
stars \citep{dub98,abadi06,rud06,cyp06,mur07,cui13}.  This is the diffuse
intracluster light (ICL) that is ubiquitous in massive galaxy
clusters.  It is comprised of stellar populations that are not
gravitationally bound to any single galaxy, but are dynamically part
of the galaxy cluster environment\footnote{While some works have differentiated between 
the spatially diffuse light as the ICL and the kinematically distinct populations as a 'diffuse stellar 
component' (DSC; \eg Dolag et al. 2010, Cui et al. 2014), we consider them both to be 
closely related}.  The ICL,
while of extremely low surface brightness, has been studied in great detail
\citep[\eg][]{jjf04a,zib05,gonz05,krick06,krick07,burke12}, and is now
known to contain a significant fraction of the total optical
luminosity of galaxy clusters ($\sim 10-30\%$; see references above);
similar fractions are predicted in theoretical studies of
galaxy cluster evolution \citep{mur04,wil04,sl05,mon06,sta06,pur07}.   

The detailed study of this extra stellar component provides a
crucial glimpse into the dynamical evolution of cluster galaxies and
the physical processes that shape them, including galaxy mergers,
tidal stripping, cluster virialization, and even the initial
conditions of cluster formation.  The ICL is also a useful probe of
the continuous evolution of the galaxy cluster in which it resides;
although the central BCG appears to have formed much of its mass
early on, the build-up of the surrounding ICL is likely a much more
gradual process, with a significant fraction of it having formed
since $z=1$ \citep[\eg][]{wil04,mur07,con07,burke12}.  The
wealth of substructure in and around nearby galaxies suggests that the
process of ICL formation continues to the present time
\citep{hos05,rud09}.

While studies of the diffuse ICL provide insight into the evolution of
galaxy clusters, the study of the {\it individual} stellar populations
that comprise this `extra' luminosity can provide even more useful
constraints on dynamical evolution.  The majority of studies of the
discrete stellar populations have been carried out through searches
(and subsequent studies) of planetary nebulae
\citep[PNe; \eg][]{jjf98,arn04,jjf04b,ag05,ger07,cr09,long13}, red giant stars
\citep{ftv98,dur02,wil07a}, and supernovae \citep{gal03,sand11}.   

\subsection{\it Globular Clusters}

Globular clusters (GCs) provide an important tool for understanding
the formation and evolution of galaxies \citep[see the reviews by][]{az98, har01, west04, bs06}.  They are found in galaxies
of all luminosities, environments, and Hubble types (the lone
exception being the lowest luminosity dwarf galaxies with $M_V\sim
-10$, which typically do not have any GCs).  GCs tend to be old
($\sim 10-12$~Gyr) and have (relatively) simple stellar populations.  Thus their 
integrated broadband optical/NIR colors can be used as a proxy for metallicity, allowing their use as
tracers of the early chemical evolution of galaxies, although the form
of their color--metallicity relation remains a critical issue
\citep[\eg][]{puzia02,peng06,yoon11,blake12,vander13}.  Furthermore, GCs
are the most luminous stellar systems within galaxies, allowing their visibility 
to greater distances than other stellar tracers -- studies of
GC systems extend to $\sim 100$ Mpc and well beyond
\citep[\eg][]{har09,west11,a-m13}.

GCs are so ubiquitous that they are useful probes of not only 
individual galaxies, but of the galactic environments in which they
reside, from sparse galaxy groups to the densest galaxy clusters.
Many studies of GC systems around
galaxies have centered on the massive elliptical galaxies in dense
galaxy clusters that have the largest GC populations
\citep[\eg][]{har86,geis96,blake97,blake99,har09,har09a,har09b,puzia14}.  However, GC systems
typically are quite extended, and thus studies of complete GC systems
around individual galaxies are difficult to carry out, especially for nearby galaxies where large area coverage is
required.  The availability of large CCD mosaics now enables more
routine studies of not only the entire GC populations around nearby
galaxies, but also {\it cluster-wide} GC populations.

The term `intergalactic tramp' was used in the 1950's
\citep{vdb56,bs58,vdb58} to describe globular clusters at large
distances from the Milky Way.  However, it was the early works by
Muzzio and collaborators \citep[][]{forte82,muz84,muz86,muz87} that 
began to explore the scenario of `GC swapping' between galaxies in
clusters.  Both \citet{muz86} and \citet{white87} concluded that some
of these GCs would become members of the cluster itself -- intracluster GCs (or IGCs).
\citet{west95} more formally developed the idea that a large IGC 
population could help explain the high GC specific frequency $S_N$
values for large ellipticals in the centers of many clusters (\eg M87
in Virgo).  N-body simulations by \citet{yb05} and \citet{by06}
suggested -- consistent with studies of intracluster stars -- that IGC
populations in rich clusters should comprise $\sim 20-40\%$ of the
total cluster GC population.

While cluster-wide GC populations are predicted to exist, their
definitive detection has been elusive, largely because of the
extremely low number densities expected (thousands of GCs spread out over an
entire galaxy cluster).  Searches for GCs using broad-band colors are
necessarily statistical in nature, due to a significant `background'
of Milky Way stars and distant galaxies.  For example,
\citet{tam06a,tam06b} found that any IGC population in their study of the Virgo
cluster core region would be at the $\Sigma_{GC} =
0.1-0.5$~arcmin$^{-2}$ level, limited by background uncertainty.
\citet{bass03} found a spatially extended excess of objects with GC-like colors in the
Fornax cluster, consistent with a cluster-wide population \citep[also
suggested by][]{grill94,k-p99,berg07}, but follow-up spectroscopy showed that
most are likely bound to NGC~1399 \citep{sch08}.

The first significant detection of a cluster-wide GC population was
that by \citet{jor03} in the center of Abell 1185, later confirmed
with color information by \citet{west11}.  Since then, small numbers
of IGCs have been confirmed in the Virgo and Fornax clusters through
the visual inspection of HST images \citep[where the GCs are resolved,
and some red giant stars visible;][]{wil07b} or through spectroscopy
\citep{firth08,sch08}.

The first wide-field surveys of sufficient photometric depth to detect
statistically significant cluster-wide GC population were carried out
in the Virgo cluster \citep{lee10} and in the Coma cluster
\citep{peng11}.  Both showed a significant number of GCs outside the
large galaxies of these clusters; a similar result was found with HST
in the much more distant cluster Abell~1689 \citep{a-m13}.  Like red
giant stars, GCs contain information about the metallicity and age of
their parent population; like planetary nebulae, GCs are useful
tracers at very low surface densities; like ultra-low surface
brightness photometry, GC surveys can reach distances far beyond that
possible via individual stars.  Perhaps most importantly, GCs can
serve as dynamical tracers throughout an entire cluster, making them
an important yet virtually untapped resource for understanding the
evolutionary states of galaxy clusters.

\subsection{\it GCs in the Virgo Cluster}

At a mean distance of 16.5 Mpc \citep{ton01,mei07}, the Virgo Cluster provides the best laboratory for studying the formation and evolution of a typical galaxy cluster.   The measured distance of the cD galaxy M87 is consistent with the cluster mean \citep{blake09}.  However, the cluster itself appears unrelaxed, and deep surface photometry of the core region around M87 by \citet{hos05} clearly shows a wealth of low surface brightness
streams and other diffuse features, indicative of a complex evolutionary history.  Dynamical studies further showed the complexity
of the M87 GC system itself \citep{cohen00,cote01,strad11,rom12,zhu14}.  The low surface
brightness features and substructure are not confined to the Virgo core, but also occur around other massive galaxies and subgroups
within Virgo \citep{jan10,hos13,paudel13}.  However, the cluster's large angular extent \citep[the photographic survey by][covered a
total of $\sim 140$ square degrees]{vcc}, makes cluster-wide ICL studies extremely difficult, particularly at resolutions that allow
detailed mapping at the galactic level.

Previous photometric and spectroscopic studies of GCs in Virgo have
covered limited areas near individual galaxies, or larger areas to
very shallow depths.  However, hints of how the large-scale
environment shapes galactic GC systems can be found in many of these
studies.  For instance, as part of the ACS Virgo Cluster Survey
\citep[ACSVCS;][]{cote04}, \citet{peng08} found that dwarf galaxies near M87 
have richer GC populations than those farther out, suggesting biased
GC formation.  On the other hand, a possible increase of $S_N$ (also
measured from ACSVCS) for giant elliptical galaxies (excluding M87)
with cluster-centric radius may be a sign of tidal stripping of the
outskirts of these galaxies \citep{coe09}.   \citet{lee10} used
SDSS photometry to study the Virgo-wide population of GCs, but only to
$i<21.7$, at which just $\sim 13\%$ of the complete GC population is 
detectable.  These authors did find a statistically significant
population of GCs throughout the inner regions of the Virgo Cluster,
suggesting a large number of IGCs.  However, the relatively low
spatial resolution (limited by the small numbers of GCs detectable at
SDSS depth) did not allow for a clear segregation into galactic and
intracluster GC populations.

Furthermore, Virgo is a less massive, less relaxed cluster than the
Coma cluster, the only other galaxy cluster in which a large-scale
($\sim1$~Mpc) study of the cluster-wide GC population has been
conducted \citep{peng11}.  Thus, an accurate census of Virgo's overall
GC population would allow for useful comparisons of the intracluster
stellar components in clusters of differing masses and evolutionary
states.  Clearly, a very deep, very wide, photometric survey of the
Virgo cluster has been badly needed.

\subsection{The Next Generation Virgo Cluster Survey (NGVS)}
    
The Next Generation Virgo Cluster Survey \citep[NGVS;][]{ngvs} is a
very deep, multi-band photometric survey covering 104 deg$^2$ of sky in the direction of the Virgo cluster.  The photometric depth of NGVS
(see next section) is sufficient to reveal the majority of the GC
population in Virgo.  Thus, it combines the depth of some previous
targeted studies \citep[\eg][]{tam06a,tam06b}, but at twice the spatial resolution, with the extremely wide field (albeit shallower) cluster-wide survey of \citet{lee10}.  The present
work focuses on the search for GCs over the entire NGVS region,
through the use of the $g'_{o}$,\gio color-magnitude diagram for
unresolved and marginally resolved sources in the vast NGVS catalog.
The exquisite data quality of the NGVS allows a thorough investigation
of the amount of background contamination, the primary source of
uncertainty in such studies.   Other papers in the NGVS series that are related to the work presented here include spatial and kinematic studies of Virgo's ultra-compact dwarf galaxies (C.Liu \etal, in preparation, H.Zhang \etal, in preparation), spectroscopy of M87's GCs (E.Peng \etal, in preparation), dynamical modeling of M87's GC system \citep{zhu14}, and the use of $u^*i'K$ imaging in the classification and detection of both stellar and galactic sources in the NGVS \citep[][]{ngvsir}.

\section{Observations and Data}

The technical details of NGVS survey itself, including the field
selection, data acquisition and data processing are presented in
\citet{ngvs}, so only a brief overview will be provided here.  The NGVS is
a large-scale survey covering the Virgo
Cluster out to one virial radius, with deep images using the CFHT MegaCam wide-field optical
imager and the $u^*g'i'z'$ filters, with partial coverage in the $r'$ band.  The MegaCam field-of-view
for a single image is roughly $0.90$ deg$^{2}$, although with the
dithering strategy employed, the total field of view for each stacked
NGVS field is $\sim 1.0$ deg$^2$.  The total NGVS survey consists of
117 separate fields, and allowing for overlap between fields, the sky
coverage of the NGVS is more than $104$ deg$^2$.  This covers Virgo
subclusters A and B out to their virial radii.  The NGVS dataset also
includes four separate control fields (hereafter labeled BKG1-4) that
are located $\sim 10^\circ$ away from the NGVS footprint; these images are
of similar depth and image quality as the rest of the NGVS images, and
are thus useful for defining the characteristics of background contamination and assessing the variations in the Galactic foreground stellar population.

Calibration of the NGVS data is done through comparison to stars from
the Sloan Digital Sky Survey \citep{sdss} through the MegaPipe
calibration pipeline \citep{gwyn08}, with rms errors of $0.01-0.02$ mag.  
Unless noted otherwise, the resulting magnitudes are on the CFHT MegaPrime photometric system.  Although the MegaPrime
$u^*g'r'i'z'$ system shares similarities with the SDSS $ugriz$ system,
there are notable differences, especially with the $u^*$ filter
\citep[\eg][]{gwyn08, ngvs}; any references to the SDSS $ugriz$ system
will be explicitly stated.

\subsection{Photometry}

Photometry of all objects on the final, stacked NGVS images was
conducted using SExtractor \citep{ba96}; the details can be found in
S. Gwyn \etal (in preparation).   We use only the NGVS $g^\prime$ and $i^\prime$ photometry, as these were the first filters for which the entire NGVS footprint was completed.    Future papers will make use of the $u^*i'K$ color-color diagram \citep[][]{ngvsir} for improved globular cluster selection in the NGVS dataset.   Here we use the NGVS $g'i'$ photometry of sources based on an aperture diameter of 8 pixels.  The spatially-dependent aperture corrections (due to small variations in the stellar point-spread-function [PSF] across the MegaCam field) were derived using samples of bright, unsaturated stars in each MegaCam field for each filter.  The individual extinctions $A_{g'}$ and $A_{i'}$ values for each object were derived from the 100$\mu m$ dust maps from
\citet{sch98}.  All magnitudes presented here have been corrected for
these extinction values.

We have considered only those objects that lie outside regions masked
by the reduction pipeline following the prescription of \citet{gwyn08}.  Those are the regions around very bright, saturated
stars that contain both saturated pixels, diffraction spikes, and elevated
signal/noise in the immediate regions, rendering the photometry of
objects located near such features suspect.  While a survey
such as the NGVS will contain many such bright stars, the small size
of the individual masked regions ($\sim 0.01$ arcmin$^2$ for the
larger ones) will have an insignificant effect on the spatial number
densities of objects over the large area being considered here; the
total area around $\sim 200$ masked regions around bright stars is
$\sim 2.0 $deg$^2$, or less than 2\% of the total field area.  For our
study, we use only those objects in the NGVS catalog with $18.5<$ \go
$<24.0$; the bright cutoff is based on the saturation level of the
brightest stars, and the faint \go cutoff is based on the approximate
location of the GC luminosity function turnover for objects in the
Virgo Cluster \citep[][see also Section 3.4]{jor07}.  This choice will
maximize the number of GCs present in our analysis, while reducing
the (increasing) contamination from unresolved background galaxies at
fainter magnitudes.

Photometric incompleteness will not be a significant issue for this
study.  For much of the NGVS field area, image crowding is minimal,
and photometric incompleteness due to sky backgrounds will be
insignificant.  Our adopted faint-end cutoffs (\go$=24$, \io$\sim23.8$)
are over 1.3 magnitudes brighter than the nominal S/N=10 point source
photometry for the entire survey \citep[$g'\sim 25.9$ and $i'\sim 25.1$;][]{ngvs}, thus we expect our photometric completeness to be close to 100\%.  However, a pair of small regions within 2' of the centers of M87 and M49 were not included in the SExtractor catalogs due to the high surface brightness background.    For M87, a supplemental photometric catalog of sources in the
central four deg$^2$ was derived from NGVS images with the light from M87 subtracted; any objects found and measured in the M87-subtracted
images were folded in to the original catalog\footnote{ A small number of objects near the center of M49 were missed in the SExtractor catalog, although a comparison of our source lists with confirmed M49 GCs 
show that a very small number (12) of the confirmed GCs were missed.    The small number of missing objects will have no effect on the results presented in this paper, which are based on the larger-scale distribution of GCs throughout the Virgo cluster.}

\subsection{Point Source Selection}

At the distance of the Virgo cluster, many (but not all) globular
clusters are expected to be point sources in the NGVS $i'$ images.  At
the faint photometric levels explored here, contamination by
background galaxies can be significant, so the careful removal of clearly resolved objects is crucial in extracting the
(comparatively) small number densities of GCs throughout the survey
area.  For a Virgo cluster distance of $16.5$ Mpc, the median seeing
of $\sim 0$\farcs$6$ for most of the NGVS $i'$ images corresponds to a
physical size of 48 pc.  Sources are expected to be marginally
resolved for sizes of at least $\sim 0.1$ FWHM \citep{har09a}, or $\sim
5$pc in our data.  With typical Virgo cluster GCs having $r_h\sim
2-4$pc \citep[\eg][]{jor09,strad11,puzia14}, we expect enough GCs to be
marginally resolved such that we need to be mindful of throwing out
significant numbers of bona-fide GCs through the use of strict
point-source selection parameters.

Point sources were selected based on the concentration
index \deli = $i'_{4} - i'_{8}$, the difference between the 4-pixel
and 8-pixel diameter $i'$ {\it aperture-corrected} magnitudes.  Only
the $i'$ magnitudes were used, as (a) the NGVS $i'$ images were taken
under the best seeing conditions, and (b) the field-to-field variance
in FWHM in the $i'$ images was much less than that for the $g'$
images.  Thus non-stellar sources will be more clearly resolved in the
$i'$ images -- tests using a similarly-defined $\Delta g'$ index shows
the \deli indication alone is an effective source discriminator.

As bright stars were used to create the aperture corrections described
in the previous section, all true point sources should have \deli$\sim
0$; resolved sources will have progressively larger (positive) values
of \deli.   This is shown in Figure 1, which illustrates \deli values for
sources in the entire NGVS catalog.

\begin{figure}
\epsscale{1.00}
\plotone{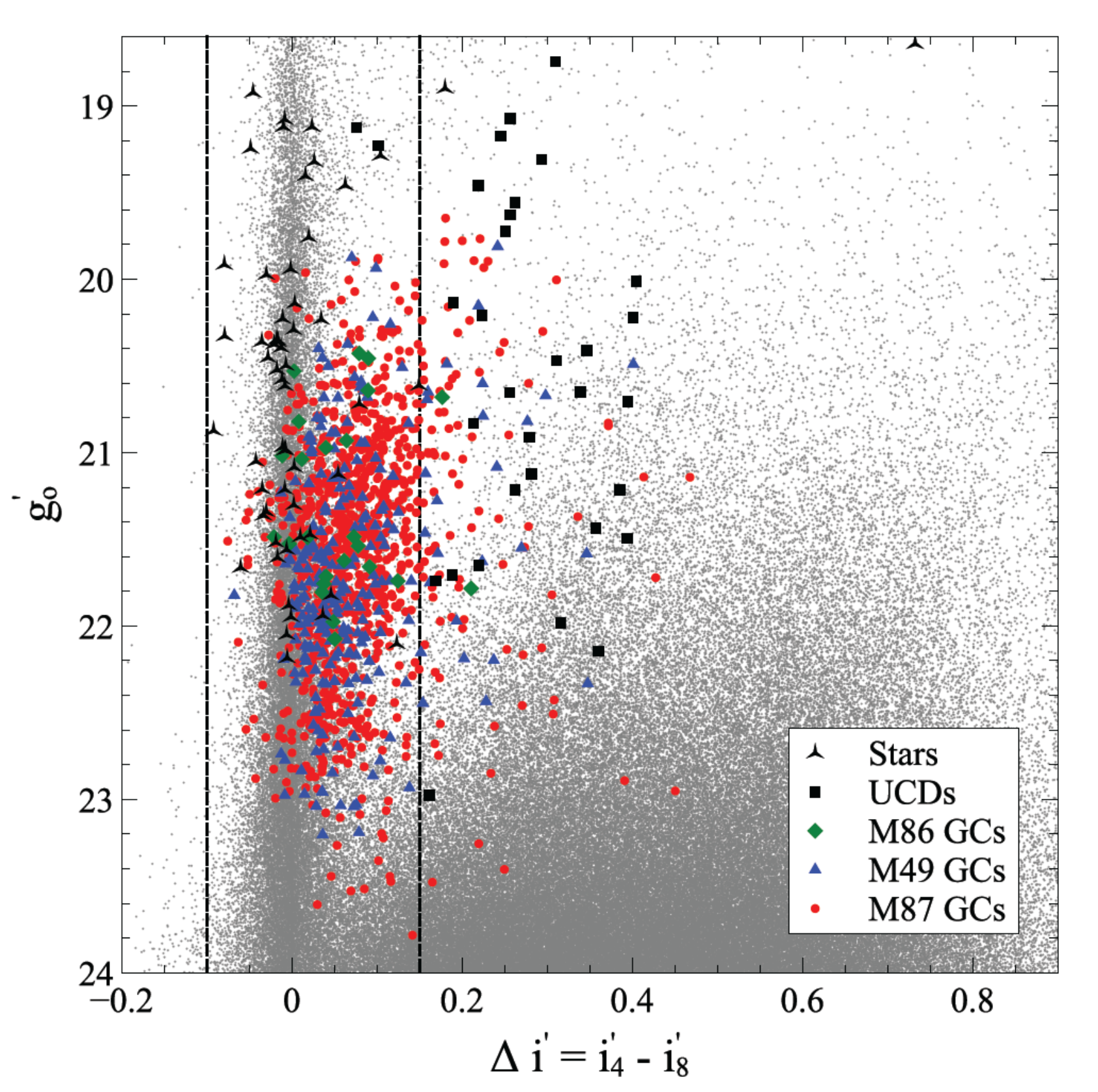}
\caption{Plot of our $i'$-band concentration index \deli $= i^{'}_{4} - i^{'}_{8}$ 
versus the total $g'$ magnitudes for sources (only 5\% of sources plotted for clarity) within the NGVS
footprint (grey points); the stellar sequence at \deli $\sim 0$ is
clearly visible.  Overplotted are NGVS-based photometry for 1216 GCs
in 3 Virgo cluster galaxies that have spectroscopically-derived
velocities consistent with Virgo cluster membership (see text for
details).  Also included are Virgo
UCD galaxies from \citet{bro11} and confirmed stars from \citet{strad11}.  Many of the confirmed
GCs lie at larger \deli values than for stellar objects, indicating
that a significant fraction of GCs are indeed marginally-resolved on
the $\sim$0\secpoint 6 seeing of our $i'$ images.  The dashed lines
show the adopted \deli values for our point source catalog.\label{fig1}}
\end{figure}

To determine the range of \deli appropriate for Virgo GCs, we have
compiled a master catalog of confirmed GCs in some of the luminous
Virgo cluster galaxies.  This catalog includes all GCs with
spectroscopically-derived velocities consistent with Virgo cluster
membership, with M87 GCs from \citet{hanes01} and
\citet{strad11}\footnote{We consider all \citet{strad11} objects classified as GC 
or as transitional objects as bona-fide GCs.}, M49 GCs from \citet{cote03}, and
M86 GCs from \citet{park12}.   We have also included 253
velocity-confirmed M87 GCs from Peng \etal (in preparation) that
are neither in the Hanes et al. nor the Strader et al. catalogs.  The final list
includes a total of 1242 GCs from the 3 galaxies combined, of which
1218 have $g'i'$ photometry from the NGVS catalog (the very few missing objects are
located either within masked regions or near the center of M49).

The \deli values for the spectroscopic GC sample are also plotted in Figure 1,
along with 56 confirmed stellar sources in the M87 GC catalog of
\citet{strad11}, and 32 Virgo cluster UCDs from \citet{bro11}.  Here
we see that a significant number of Virgo GCs are
marginally resolved, with \deli values systematically larger than that
of the mean stellar sequence at \deli=0 .  A few GCs (and most of the
cataloged UCDs) are well resolved in our data, with larger \deli
values.  As our goal is to extract as many GCs from the NGVS data
without allowing many resolved contaminants into our sample, we have chosen our source selection criterion as those objects with $-0.10 <$
\deli $<+0.15$, based on visual inspection of Figure 1.  This criterion contains 1094
confirmed GCs, or $90$\% of the complete sample.  Hereafter, all
objects that pass the \deli criterion will be referred to as point
sources, although we re-iterate that some marginally-resolved sources
are included in this subsample.  Our 'point source' sample -- $18.5< $
\go $<24.0$ and $-0.10<$ \deli $<0.15$ -- for the entire NGVS region
(excluding the four background fields) contains 792790 objects.

\subsection{Globular Cluster Color Selection}

At the magnitudes and colors expected for Virgo cluster GCs, there
will be a significant contribution from both foreground
Milky Way stars and from unresolved background galaxies.  To select
candidate GCs, we make use of specific regions within the $g'_{o}$, \gio
color-magnitude diagram (CMD).

The left-hand panel of Figure 2 shows a smoothed Hess diagram for all
of the objects in our point source catalog.  Each object in the CMD
has been Gaussian-smoothed {\bf by the} photometric errors.  The Hess diagram is comprised of many
different populations, but is dominated by main-sequence turnoff
(MSTO) stars with \gio $\sim 0.4$ from both the intervening Milky
Way halo (and its substructures) and many nearby, low
luminosity MW disk stars visible at
\gio $\sim 2.0-2.5$.  At the faintest magnitudes there are many
unresolved background galaxies at all colors.  The locus of point
sources at \gio $\sim 0.6-0.8$ is the superposition of two distinct
populations -- those objects with $18.5 <$ \go $< 21.0$ are largely
red-giant-branch (RGB) stars from the Sagittarius dwarf galaxy (which
passes through the center of the NGVS region; see Section 3.1.1),
while fainter objects are GCs at the distance of the Virgo cluster.

\begin{figure*}
\includegraphics[scale=0.57]{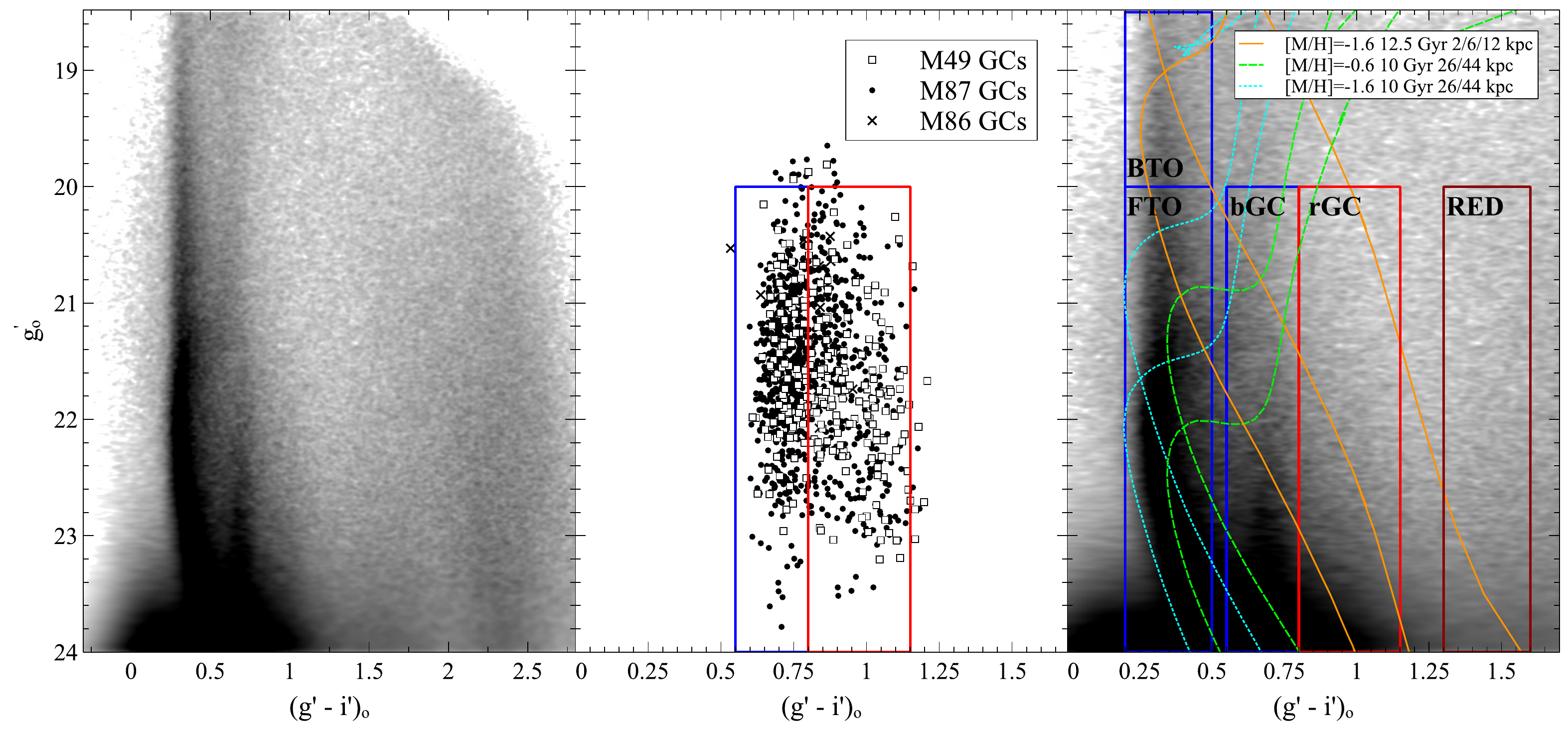} 
\caption{(left) \go, \gio
 Hess diagram for all point sources that 
satisfy the \deli criterion (Figure 1) and have $18.5< $\go $< 24.0$.  The colors and magnitudes for
all sources have been smoothed by a value of the combined
photometric errors and a constant value.  (center) \go, \gio 
color-magnitude diagram for radial velocity-confirmed Virgo cluster GCs; the colored boundaries show the adopted color and magnitude criteria used for blue GCs (bGCs) and the red GCs (rGCs).   (right) The 
\go, \gio Hess diagram for point sources overplotted with Marigo \etal (2008) isochrones with representative 
age/metallicity values appropriate for stars in the Milky Way halo at
distances of 2-16 kpc (orange lines), and for stars expected to be in
the Sagittarius dwarf galaxy (cyan and green lines) incorporating a
spread in both the distance and metallicity; see text for details.   The boxed regions are the locations of the 
(from top left to bottom right) bright MW halo turnoff stars (BTO),
faint [Sgr] turnoff stars (FTO), the blue globular cluster region
(bGC), the adjoining red globular cluster region (rGC) and the red
halo star populations (RED).\label{fig2}}
\end{figure*}

To determine the appropriate \gio color range for Virgo GCs, the
center panel of Figure 2 shows NGVS $g'i'$ photometry of the
radial velocity-confirmed GC sample (see previous section).  Most GCs have
$0.55<$ \gio $<1.15$ and \go$>20.0$, which we adopt as our color and
magnitude criteria for GC selection over the entire NGVS dataset.
1055 (or 87\%) of the 1218 velocity-confirmed GCs in our sample
satisfy all three criteria for selection as GCs.

As the range of \gio colors are indicative of the range
of metallicities of the (assumed old) GCs, we have further divided our
color criteria into a blue metal-poor GC population (bGC) and a red
metal-rich GC population (rGC), with the dividing color at \go
$=0.80$.  This color corresponds to a metallicity [Fe/H]$\sim -1.05$
based on the $(g-i)_o-$[Fe/H] relations from \citet{lee10} and
\citet{sinn10}\footnote{We first converted our
MegaCam \gio color to the SDSS $(g-i)_o$ color using the relations in
Gwyn (2008).}.  It is important to point out that this color choice was
based not on a preferred metallicity, but on the most appropriate
color above which very few Sgr stars will appear, aiding in our
definition of the appropriate background (control) samples, described
below.

\subsection{Foreground/Background Contamination}

The huge areal coverage of the NGVS and the use of a single broadband color to select GCs also means 
a large number of contaminating stars and galaxies is present\footnote{Use of multiple colors will reduce the
background contamination \citep[\eg][]{rz04,tam06a,kim13}; future
studies will make use of the multi-color NGVS data
\citep[\eg][]{ngvsir}.}.  The presence of numerous
spatially distinct halo populations at different distances along the
line-of-sight to the Virgo cluster makes the stellar contamination
highly spatially dependent; such features include (but are not limited
to) the leading arm of the Sagittarius dwarf galaxy \citep{bel06}, the
Virgo Overdensity \citep[VOD;][]{juric08} and the related Virgo
Stellar Stream \cite[VSS;][]{duff06}.  The structure of these features
based on NGVS observations will be discussed in future papers in this
series.

To illustrate this, the right-hand panel of Figure 2 shows isochrones
approximating the expected stellar populations within the NGVS data.  One of the
leading arms of the Sagittarius dwarf galaxy \citep['Stream A' from][]{bel06} passes through the NGVS field.  Moreover, the mean
heliocentric distance of this stream changes significantly over the
NGVS region, from $d\sim 26$ kpc for our background B3 field to $\sim
44$ kpc in our background B4 field, with a range of $29-40$ kpc for the
NGVS science fields \citep{bel06,nied10}.  We have over-plotted
isochrones that span the ranges of both distance and metallicity of
Sgr stars; we used the \citet{bress12} isochrones of ages 10 Gyr, with
metallicities of [M/H]$\sim-1.6$ and $-0.6$\footnote{Our results do
not, however, depend strongly on the representative values used here.}.
Most Sgr main sequence and subgiant stars will have \go $>20$ with
colors \gio $< 0.8$.  Similarly, any MW halo main sequence stars at
$d< 20$ kpc appear at brighter magnitudes;
these are indicated by the (old; 12.5 Gyr) Bressan et al. isochrones
with [M/H]$=-1.6$, consistent with the mean metallicity of MW halo
stars \citep[][]{rn91,an13}.

To quantify the spatially variable contamination from
both halo sources and background galaxies, we have defined sections of the $g'_{o}$,
\gio CMD (shown in Figure 2) to make the cleanest
possible distinction between the many different populations present in
the CMD:

\begin{itemize}

\item BTO: $18.5<$ \go$< 20$, $0.2< $\gio$<0.5$; largely contains 'bright' turnoff stars from inner MW halo populations

\item FTO: $20<$ \go$< 24$, $0.2< $\gio$<0.5$; largely contains 'faint' turnoff stars from both the Sgr dwarf and any MW halo stars at similar distances

\item bGC: $20<$ \go$< 24$, $0.55< $\gio$<0.80$; samples the Virgo blue GC population, as well as Sgr main sequence, subgiant and RGB stars, and inner halo main sequence stars

\item rGC: $20<$ \go $< 24$, $0.80< $\gio $<1.15$; samples the Virgo red GC population, as well as inner halo main sequence stars and faint MW disk stars

\item RED: $20<$ \go $< 24$, $1.30< $\gio $<1.50$; samples both inner halo main sequence stars, and any faint MW disk stars\footnote{We avoid the red M dwarf sequence at \gio $> 2$, which will have little predictive power as control samples for the GC populations we are considering here.}.

\end{itemize}

From Figure 2 we see that much of the contamination for the blue GC
(bGC) region of the CMD will come from Sgr main sequence and subgiant
stars, with contributions from closer halo stars (the faint turnoff [FTO] and bright turnoff [BTO]
regions, respectively), while the primary contamination in the red GC
(rGC) region is largely from halo stars and MW disk stars, with little
contribution from the Sgr dwarf.  Unresolved background
galaxies contribute to all of the fainter regions.

To define the backgrounds for each of the bGC and rGC populations, we
used a combination of regions both internal and external to the
NGVS.   These regions are shown in Figure 3, overlaid on a map of the
point sources over the entire NGVS survey region (including the four 
background fields B1-B4).  However, as the background fields sample
parts of the MW halo that differ from that of the NGVS itself (indeed,
one can see significant changes in stellar density in fields B1 and B2
versus B3 and B4), we have also defined 10 additional regions in the
outskirts of the NGVS field (a) where no large galaxies are present
and (b) far from the central regions of either of Virgo's two dominant subclusters. 

\begin{figure}
\epsscale{1.10}
\plotone{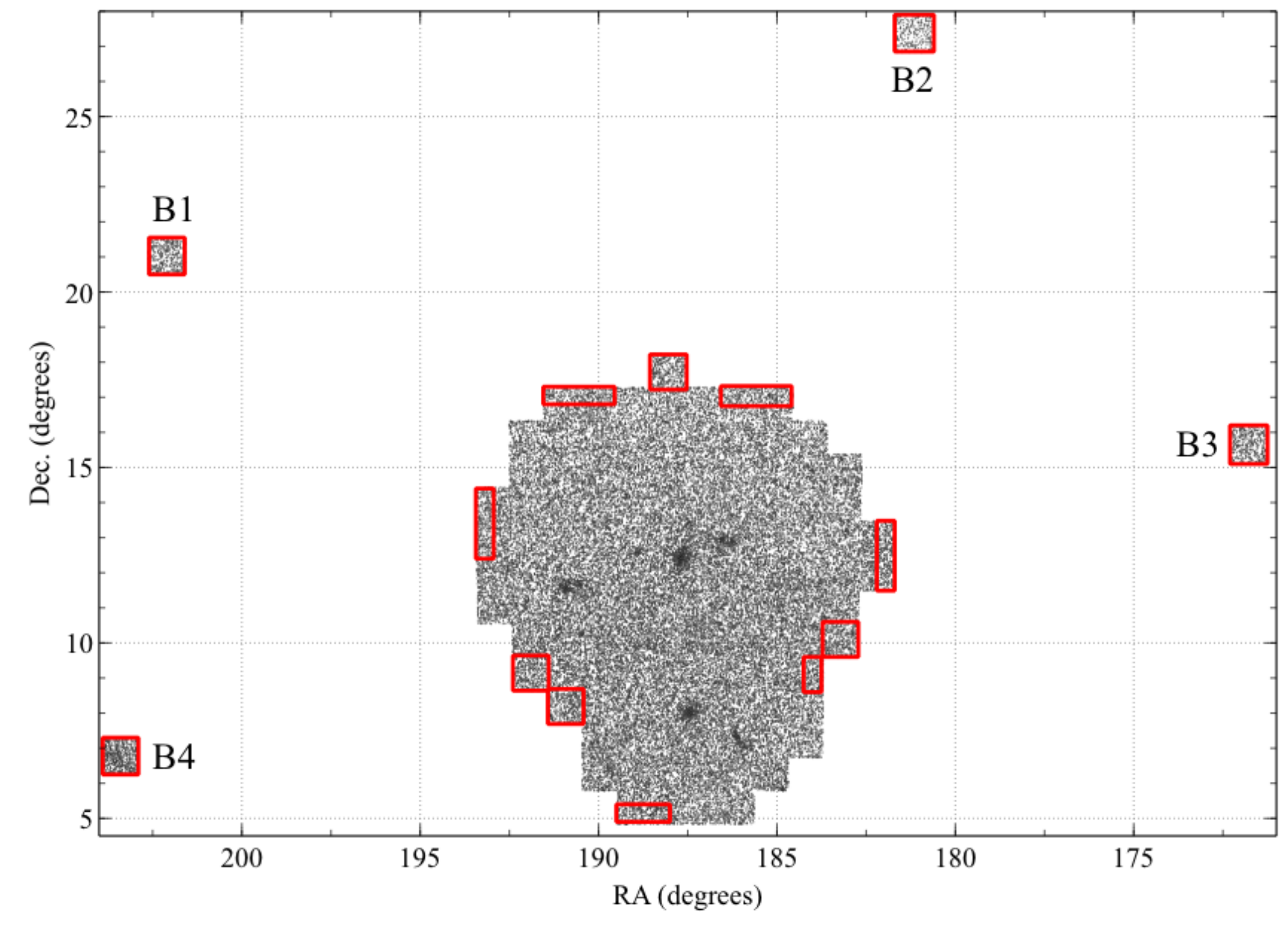} 
\caption{Spatial distribution of point sources with $18.5 < g'_o < 24.0$ within the 
NGVS survey area (center) and the background fields B1-B4.  Dense
regions around the major Virgo galaxies are apparent; however, a large
fraction of the points far from the galaxies are either foreground
Milky Way stars or are unresolved background galaxies.  The boxes (colored red in the online version)
indicate the locations of the background regions used for estimation
of background contamination in our point source catalog. \label{fig3}}
\end{figure}

For each of the regions illustrated in Figure 3, we have derived the
surface number density $\Sigma$ (in units of arcmin$^{-2}$) of point sources
through star counts in the CMD regions defined above.  We then compare
the number densities of contaminating point source objects in the bGC
and rGC regions with the densities of objects sampling the other
regions of the CMD; these are plotted in Figure 4.  The approximately
linear behavior between the number densities of GC 'contaminants' and
that of the proxy regions of the CMDs shows our simple background
models are reasonable. Furthermore, that the number densities of the
control fields within the NGVS region are consistent (within the
errors) with the 4 background fields indicate our choice of background
fields was also reasonable.  Put another way, the 10 intra-NGVS regions
do not show any statistically significant excess of objects (over the
four background fields) in the bGC and rGC regions that could be
construed as a genuine GC population in the very outskirts of the
Virgo cluster.
 
 \begin{figure} 
\epsscale{1.30} 
\plotone{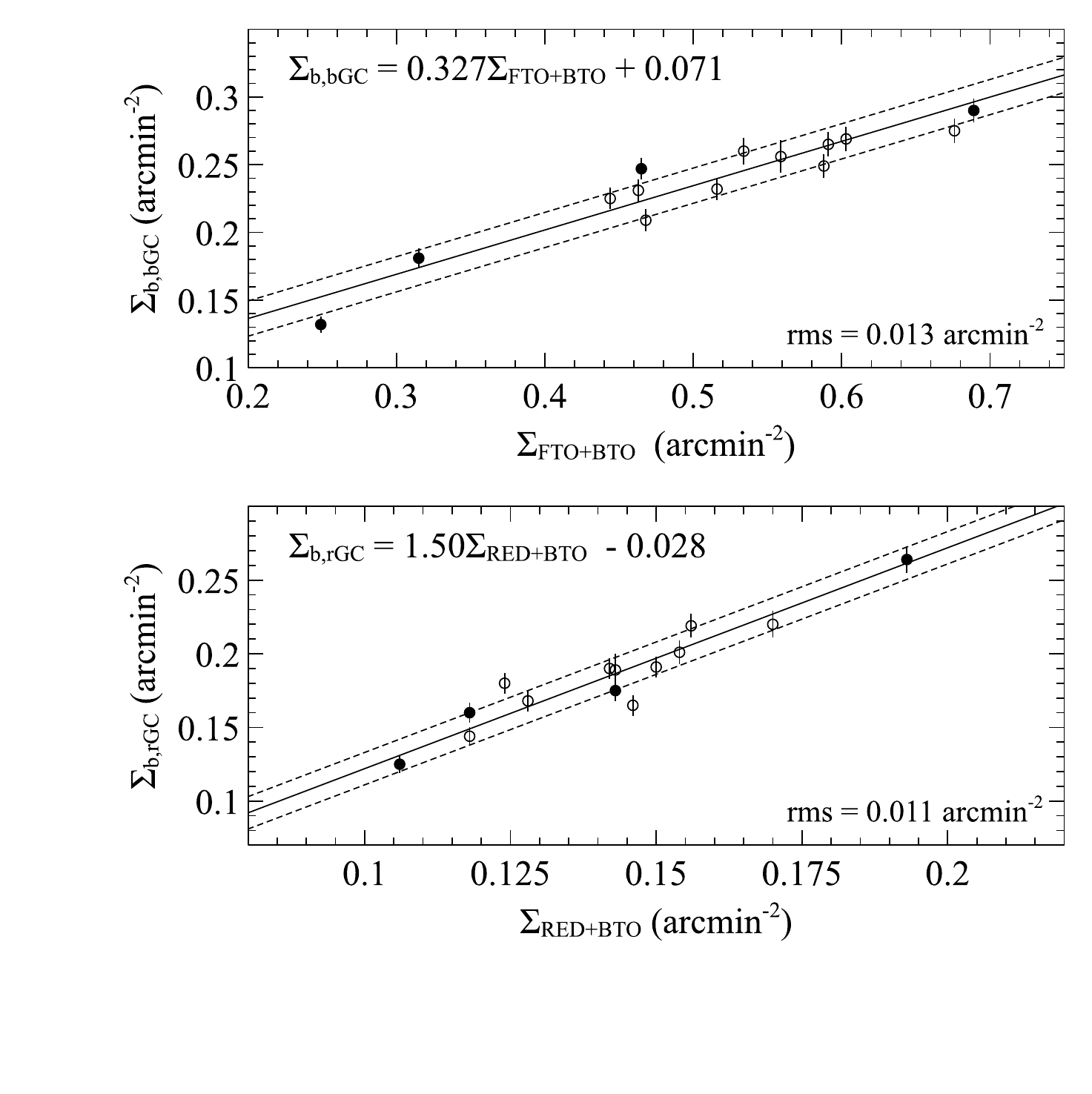}
\caption{Measured stellar surface densities (in arcmin$^{-2}$) for the expected backgrounds
for each of the blue and red GC populations, based on raw number counts in the 4 control
fields (filled circles) and that of 10 regions on the outskirts of the
NGVS field (open circles).  (top panel) The surface density of objects with colors of blue GCs as a function of the combined
number density of 
Sgr turnoff stars (FTO) and halo turnoff stars (BTO). (bottom
panel) The surface density of objects with colors and
magnitudes consistent with red GCs as a function of the combined
number density of halo turnoff stars (BTO) and halo/disk stars (RED).  In both panels
the best fitting line (and uncertainty) is shown.\label{fig4}} 
\end{figure}

Figure 4 also includes the best-fit linear fits \citep[using
orthogonal weighted regression for points with errors in both
axes;][]{fb92} to the number densities from all 14 control fields.
The mean rms errors in the $\Sigma_{b,bGC}$ relation is 0.013
arcmin$^{-2}$, and 0.011 arcmin$^{-2}$ for the $\Sigma_{b,rGC}$
relation.  These values represent the minimum possible uncertainties
in the estimate of the background populations, and lower limits on the
number densities of GCs that could possibly be extracted from our
dataset.

\section{Analysis and Results}
\subsection{Number Density Maps}

Due to the small number densities of GCs expected over such a wide
area, and the significant number of `background' objects, it is
necessary to create smoothed 2D surface density maps of the various
stellar/extragalactic/GC populations present in our data.

All spatial maps presented here were created by replacing the position
($\alpha$,$\delta$) of each point source from the different regions of
the CMD in Figure 2 with a circular Gaussian kernel with either a
constant or variable FWHM (see below).  The results were mapped onto a
full mosaic image of the NGVS field of view with a pixel scale of
$\Delta \alpha= \Delta \delta = 0.238\arcmin$.  To account for the
small variations in the true areal coverage of each pixel over the
large field of the NGVS, a pixel-area map was constructed in order to
convert the image from units of pixel$^{-2}$ to units of
arcmin$^{-2}$.  We stress that all maps presented here are based on
the {\it observed} number densities of point sources with $g'_o<24$;
unless explicitly stated, conversion of our \gc values to a total GC
number density $\Sigma_{GC,tot}$ requires a correction factor of
$2.05\pm 0.22$.  This factor is based on the detection of the top $56\pm6\%$ of the GCLF (see Section 3.4), with an additional correction for the 
$13\%$ of GCs that are removed by the point source classification described above.

While such smoothing of the data will cause some information to be
extended outside the NGVS boundary, such losses are small and will not
significantly affect our presented results.  Similarly, the number of
sources removed in masked regions of the NGVS will also not affect our
results in any meaningful way.

\subsubsection{Adaptive Smoothing}

Due to the large dynamic range of observed number densities $\Sigma_{GC}$
expected in our data -- from the many GCs near the M87 core to any
sparse IGC population -- we first applied an adaptive-smoothing
approach to the data.  We created four point source lists based on the
various regions of the CMD shown in Figure 2: the bGC region, the
rGC region, the combined BFTO and FTO regions (background for the bGC
population), and the combined BTO and RED regions (background for the
rGC population).  The position of each source was replaced by a
circular Gaussian with $\sigma = 3.5r_n$ (Gaussian FWHM=$8.26r_n$),
where $r_n$ is the radial distance from each point to its $n-$th
nearest neighbor.  To avoid any possible artifacts that might result
from the addition/subtraction of images with very different smoothing
values, the value of $n$ for each list was chosen such that the peak of the $r_n$ distribution for the lowest density regions were matched at a value $r_n \sim
0.065\degr$ (FWHM$=32\arcmin$), which approximates the largest
smoothing value used in the next section.  We have adopted the
following for each of the four maps : bGC:$r_{14}$, rGC:$r_{10}$,
BTO+FTO:$r_{26}$ and BTO+RED:$r_8$.

The resulting adaptive-smoothed maps for each of the bGC and rGC
populations are shown in Figures 5 and 6, respectively.  The left-hand
panels of both figures show the smoothed map for all point sources
with colors and magnitudes lying in the bGC/rGC regions of the CMD --
the concentrations of sources around the many large galaxies in Virgo
are readily visible.  The center panel of both figures shows the
predicted background map created through the appropriate map (BTO+FTO
and BTO+RED, respectively, for the bGC and rGC maps) and applying the
$\Sigma_{b,bGC}$ and $\Sigma_{b,rGC}$ relations from Figure 4.  Both
background maps show spatial variations, showing that adoption of a
constant background value is insufficient.

\begin{figure*} 
\epsscale{1.10} 
\plotone{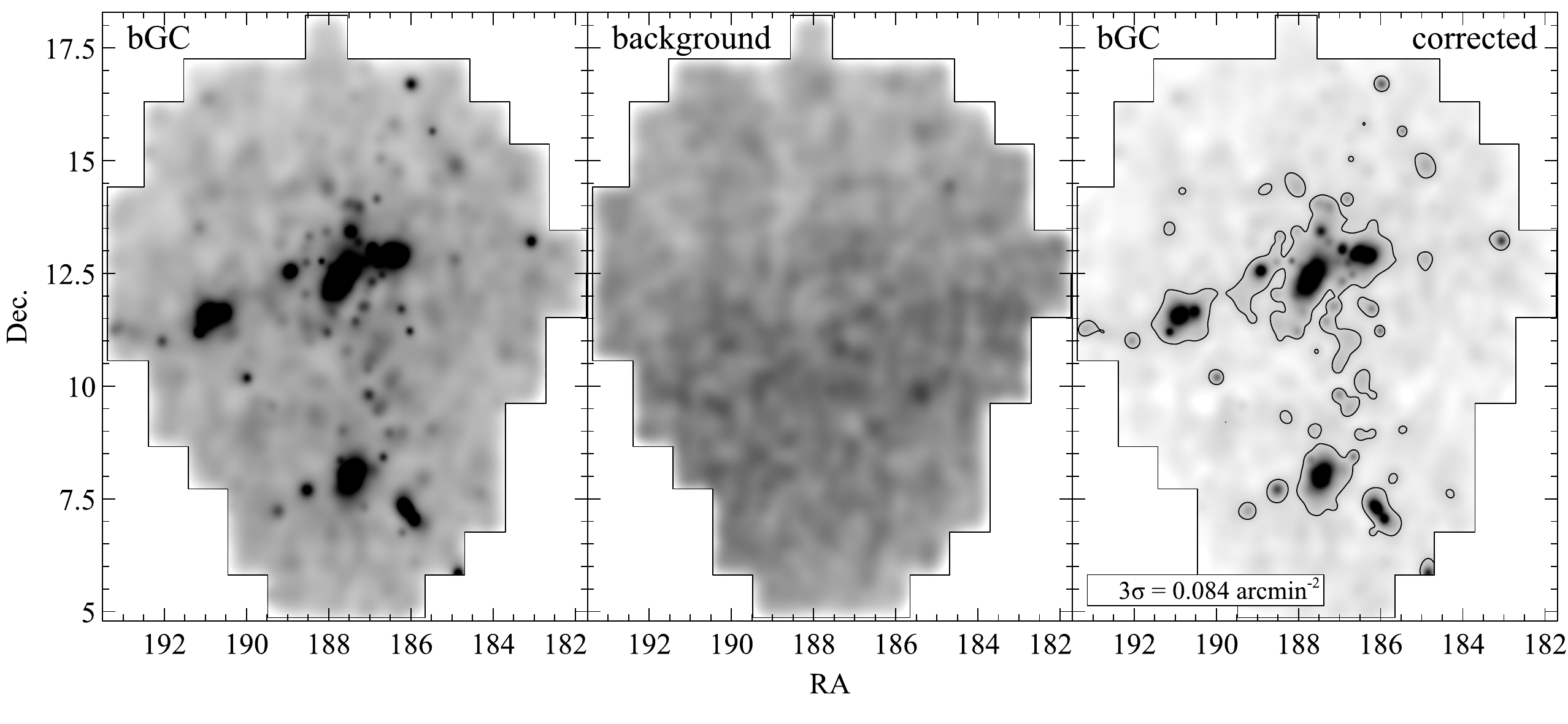}
\caption{Adaptive-smoothed bGC surface density maps:   (left) 'raw' smoothed map for for point sources with colors and magnitudes consistent with being blue globular clusters (bGCs) in the Virgo cluster. (center) model map for the expected bGC background, constructed through a (scaled) smoothed map of the combined BTO and FTO point sources.   The leading arm of the Sgr dwarf can be seen as a positive enhancement through the center of the NGVS region. (note: grayscale stretch in this image different from left-hand image in order to enhance density variations)  (right) background-subtracted map showing the surface density map of blue GCs in the Virgo cluster.   The single contour indicates the 3$\sigma$ level.
 \label{fig5}}
\end{figure*} 

\begin{figure*} 
\epsscale{1.10} 
\plotone{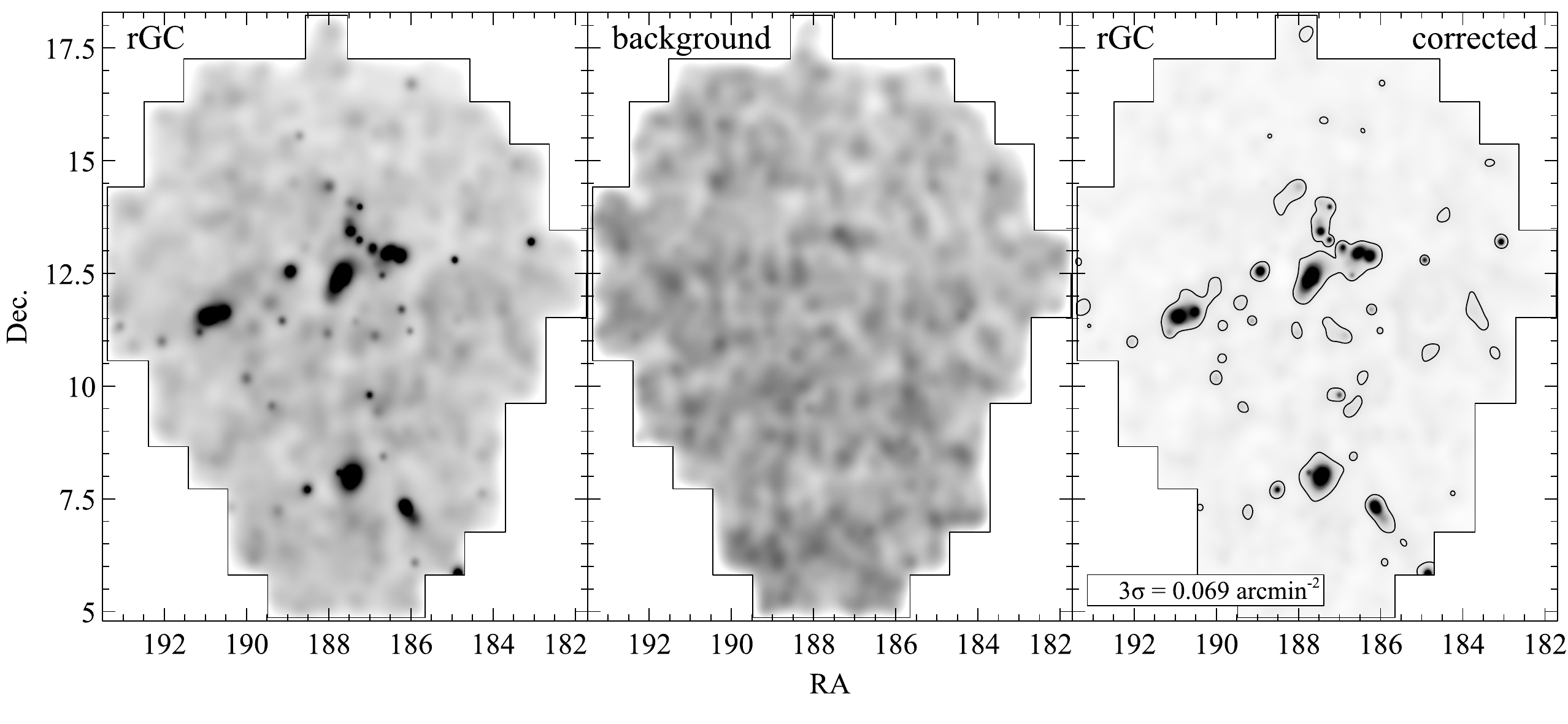}
\caption{Similar to Figure 5, but showing the adaptive-smoothed rGC surface density maps : (left) 'raw' map 
point sources with colors and magnitudes consistent with being red globular clusters (rGCs) in the Virgo cluster. (center) model map for the expected rGC background, constructed through a (scaled) smoothed map of the combined BTO and RED point sources.   (note: greyscale stretch in this image different from left-hand image in order to enhance density variations)  (right) background-subtracted map showing the surface density map of red GCs in the Virgo cluster.   The single contour indicates the 3$\sigma$ level.
 \label{fig6}} 
\end{figure*} 

The bGC background map in Figure 5 merits special mention, as the
smoothed BTO+FTO map (largely a map of main sequence turnoff stars in
the Milky Way halo) shows a distinct positive density enhancement from
$\alpha\sim 183\degr$, $\delta \sim +11\degr$, through the center of
the NGVS field, to $\alpha\sim 191\degr$, $\delta \sim +8\degr$.  This
feature is Stream A of the leading arm of the Sgr dwarf galaxy.
While a full discussion of this and other Milky Way features in the
NGVS data are left for future papers in this series, it is clear that
the entire width of this stream appears in our data.  While density
variations in the rGC background map in Figure 6 are less apparent,
there is a slight increase in the number density of stellar objects
towards lower $\delta$.  This is due to the presence of the VOD, which covers over $\sim 2000$ square degrees and
whose stars have distances of $d=6-20$kpc \citep[see][and references
therein]{bon12}.  The spatial peak of the VOD is located outside the
NGVS region towards the South; this feature is the dominant background
for the (shallower) study of Virgo's GC populations by \citet{lee10}.

The right-hand panels in Figures 5 and 6 shows the resulting
background-subtracted number density maps of the blue GC and red GC
populations in the Virgo cluster.  The single contour in these plots
is the 3$\sigma$ level; the background noise $\sigma$ for these maps
(and others) was estimated through computing the standard deviation of
the surface densities within six 300$\times$300 pixel ($\sim 71\arcmin
\times 71\arcmin$) regions in the outer regions of the map (far from
any obvious excesses).  These regions are large enough that we are
sampling true variations in $\sigma$ for all smoothing values applied.
The noise in each map is the combination of the median $\sigma$ for
the six regions and the background rms errors.

The background-subtracted rGC and bGC maps are markedly different from each other.     We illustrate this in Figure 7, which shows the 
final, background-corrected bGC and rGC maps, as well as the combined GC map (the sum of the bGC and rGC maps).    Virgo's blue GC
population appears to be more spatially extended than the red GC
population (at the $3\sigma$ background level).  A more extended blue GC distribution is also found for GC systems belonging to 
individual galaxies \citep[\eg][]{geis96,lee98,bass06,har09b,kar14} and in
the Virgo cluster from \citet{lee10}.     

However, it is possible that the spatial differences could be due to the fact that the contour levels shown in the figures are drawn at $\sigma$ levels that are relative to the background, rather than to the density level of the GC population itself. To foster a more direct comparison between the two populations, in Figure 7 we plot bCG contours drawn at the same density level as the rGC contours.   The $3\sigma$ level in our rGC map corresponds to a level $2.3\times$ larger than the mean observed rGC density  over the entire NGVS.   The same fractional level in the bGC map lies at an observed \bgc$=0.128$ arcmin$^{-2}$, or $4.5\sigma$.   This comparison shows that the bGC population is still more extended than that of the rGC population around the larger galaxies/subclumps in Virgo, but is particularly evident in the Virgo A=M87 region.  This is the clearest
demonstration to date that the difference in concentration between the red and blue GC populations is generic to all GC systems in all
galaxies, being readily apparent over the entire Virgo cluster.   We defer further discussion of specific features on these maps to Section 3.2.

\begin{figure*} 
\epsscale{1.20} 
\plotone{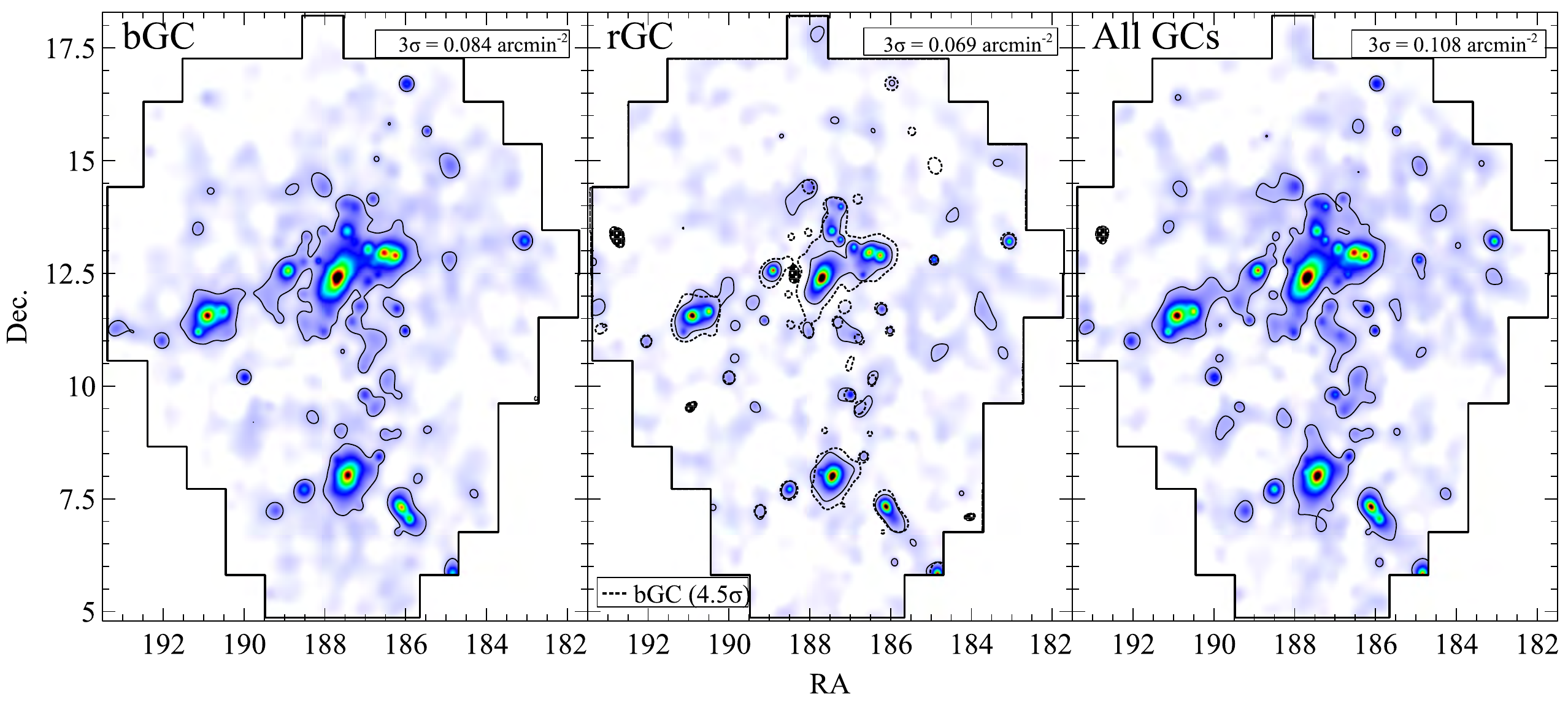}
\caption{Adaptive-smoothed surface density maps of the GC populations in the Virgo cluster (left) blue GCs, (center) red GCs, and (right) all GCs.   In each map the 3$\sigma$ contours are shown as solid lines.   Hatched regions denote the $-3\sigma$ contours in the center and right panels.    For the center panel, the dashed line represents the surface density map of the bGC population at the same density fraction as shown for the rGC population; see text for details.
 \label{fig7}} 
\end{figure*} 

As a test of our derived background $\sigma$ values for each map in
Figure 7, we also plotted the $-3\sigma$ contours as hatched regions.
There are 3 areas of the rGC map that show $3\sigma$ 'holes',
including a feature $0\fdg7$ east of M87.  All of these regions,
however, are areas where (a) the surface densities of objects in the
rGC region of the CMD show depressed values and (b) where there is an
observed enhancement in the background.   The negative feature
near the eastern edge of the NGVS field at $\alpha\sim 192\fdg8$,
$\delta\sim +13\fdg3$ coincides with the location of the galaxy
cluster Abell 1627, but it is unclear if this is the reason for the
enhanced background.  The locations of 15 other galaxy clusters in the
\citet{abell89} catalog that lie within the NGVS do not correspond
to {\it any} features -- positive or negative -- in either of the
background maps.  That such 'holes' exist on our map does indicate
that the $\sigma_{rGC}$ value may be a slight underestimate.  For
consistency we continue to use the derived $\sigma_{bGC}$ and
$\sigma_{rGC}$ values, and urge the usual caution when interpreting
3$\sigma$ features.  Similar caution is urged near the NGVS boundary
in the maps, where artifacts can be introduced from the smoothing of
objects outside (but not inside) the NGVS boundary.

\subsubsection{Constant Smoothing}

For comparison purposes, we have also created smoothed GC number
density maps using circular Gaussian kernels with a fixed FWHM$=10'$
and FWHM$=30'$, corresponding to physical scales of 48 kpc and 144
kpc, respectively.  These maps were created (and the noise $\sigma$
derived) in the same way as the adaptive-smoothed maps.  The
FWHM$=10'$ map (left-hand panel of Figure 8) puts emphasis on any
smaller-scale structures in the GC maps, at the expense of a higher
background noise.  The center panel of Figure 8 shows the more heavily
smoothed FWHM$=30'$ map, which is useful for probing the much
lower GC number densities in the regions between the galaxies.
Furthermore, as the FWHM$=30'$ map is similar in resolution (and
background noise level) of the \citet{lee10} map of Virgo's GC
population, this map will be used to foster a useful comparison with
that work.  

\begin{figure*} 
\epsscale{1.20} 
\plotone{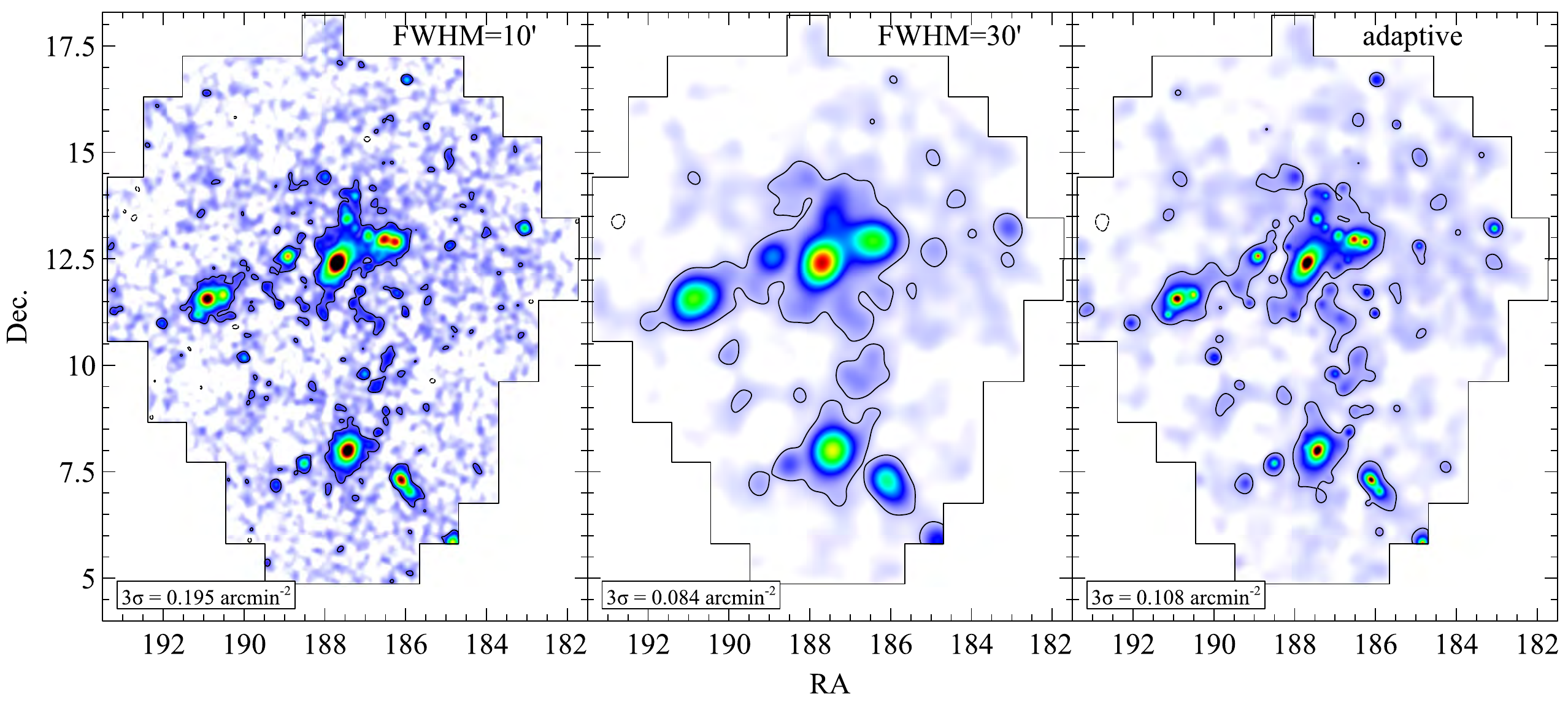}
\caption{Surface density maps of the observed GC (bGC+rGC) populations in the Virgo cluster (left) map smoothed with a Gaussian kernel with FWHM$=10'$ (center) GC map smoothed with a Gaussian kernel with FWHM$=30'$ (right) GC map smoothed with a varying (adaptive) FWHM.   In each map the 3$\sigma$ contours are shown as solid lines.   Dashed lines show the $-3\sigma$ contours.
 \label{fig8}} 
\end{figure*} 

All three maps show (at different levels of depth and
resolution) the cluster-wide distribution of GCs in the Virgo cluster, which shows 
a wealth of substructure, from the large GC populations around the luminous Virgo galaxies to
the sparse outer regions.  The adaptive-smoothed map will be used for
much of the following analyses, as it shows both fine resolution in
the cores of large galaxies (with smoothing values FWHM$\sim
3\arcmin-5\arcmin$), to much larger smoothing values (typical FWHM$\sim
20\arcmin-40\arcmin$) in the less dense regions of the NGVS.  The constant
smoothed maps will be used to test the validity of features detected
in, and the number density profiles \gc derived from, the adaptive-smoothed 
maps.  We stress that all of the features indicated in the
following sections are not artifacts of the adaptive-smoothing
process -- the same features are also visible in at least one of the
constant FWHM maps.

\subsection{The Spatial Distribution of Virgo's GC Populations}

Figure 9 shows the final, adaptive-smoothed map of Virgo's GC
(bGC+rGC) population, overlaid with the locations of the 1467 Virgo
cluster member galaxies from the VCC that lie within the NGVS
footprint.  As expected, the majority of GC features above
\gc$=0.108$ arcmin$^{-2} =3\sigma$ are spatially co-incident with
Virgo galaxies.  In particular, the largest/densest concentrations of
GCs are located in the regions surrounding the four largest substructures
of the Virgo cluster; Virgo subcluster A (centered on M87), subcluster
B (centered on M49), subcluster C (centered on M60; center left), and
the (infalling?) subcluster centered on M86, to the right of M87.  The figure also
indicates the locations of the brighter Virgo galaxies ($B_T < 16$).
What may seem surprising is the large number of bright VCC galaxies
that do {\it not} have GC peaks in the map.  Due to the (necessarily)
large smoothing values applied, any GC systems with small GC
populations (many dwarf galaxies), or low density GC systems (\eg many
late-type galaxies) will be smoothed below the $3\sigma$ contour.
Although the GC systems of many spirals are smoothed to below the
$3\sigma$ level in the number density maps, their numbers are included
when computing the total number of GCs in the Virgo cluster in Section
3.4.

\begin{figure*} 
\epsscale{1.00} 
\plotone{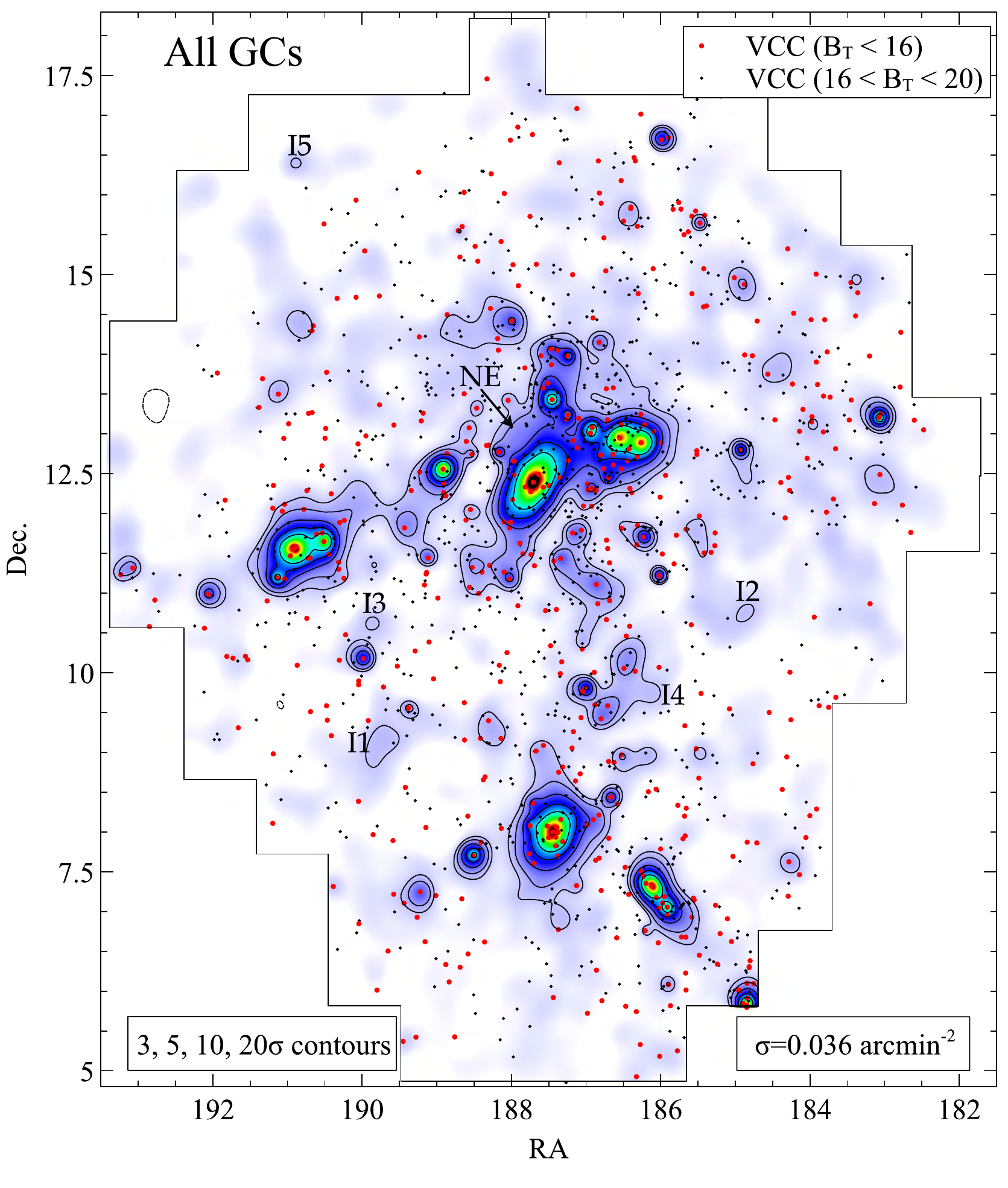}
\caption{Adaptive smoothed GC number density map of the Virgo Cluster, with the noted contours of the observed number densities \gc.    The filled circles (red in the online version of the Figure) denote the positions of Virgo member galaxies from the VCC \citep{vcc} that lie within the NGVS boundary and have $B_T < 16$; the black dots represent VCC galaxies with $16 < B_T <20$.   I1-I5 denote map features where no VCC galaxies are present, and 'NE' refers to a possible excess of GCs near M87; see text for details. \label{fig9}} 
\end{figure*} 

Figure 10 shows the location of all VCC galaxies, brighter than B=16,
that are confirmed or probable cluster members located
inside the NGVS footprint. To provide a first glimpse into the
connection between the GC distribution and galaxy morphology,
galaxies have been divided into seven broad categories based on their
appearance (i.e., isophotal structure) at optical (NGVS)
wavelengths. Full details on this scheme, including classifications
for all Virgo galaxies within the NGVS, will be presented in a future
paper in this series. Briefly, the seven categories --- E, ES, EI, S,
SI, I and T+PM --- refer to the global structure of the galaxy: i.e.,
E galaxies have smooth and regular isophotes that are nearly
elliptical in shape, S galaxies show unambiguous spiral arm patterns,
and I galaxies have a more irregular appearance. The ES, EI and SI
categories include (transitional) galaxies that cannot be assigned
uniquely to the E, S or I categories (i.e., nearly elliptical
galaxies with irregular, large-scale star formation superimposed would
be classified as type EI).  The seventh category, T+PM, refers to
galaxies that are experiencing strong tidal distortion or show
evidence for post-merger evolution.

\begin{figure*} 
\epsscale{0.80} 
\plotone{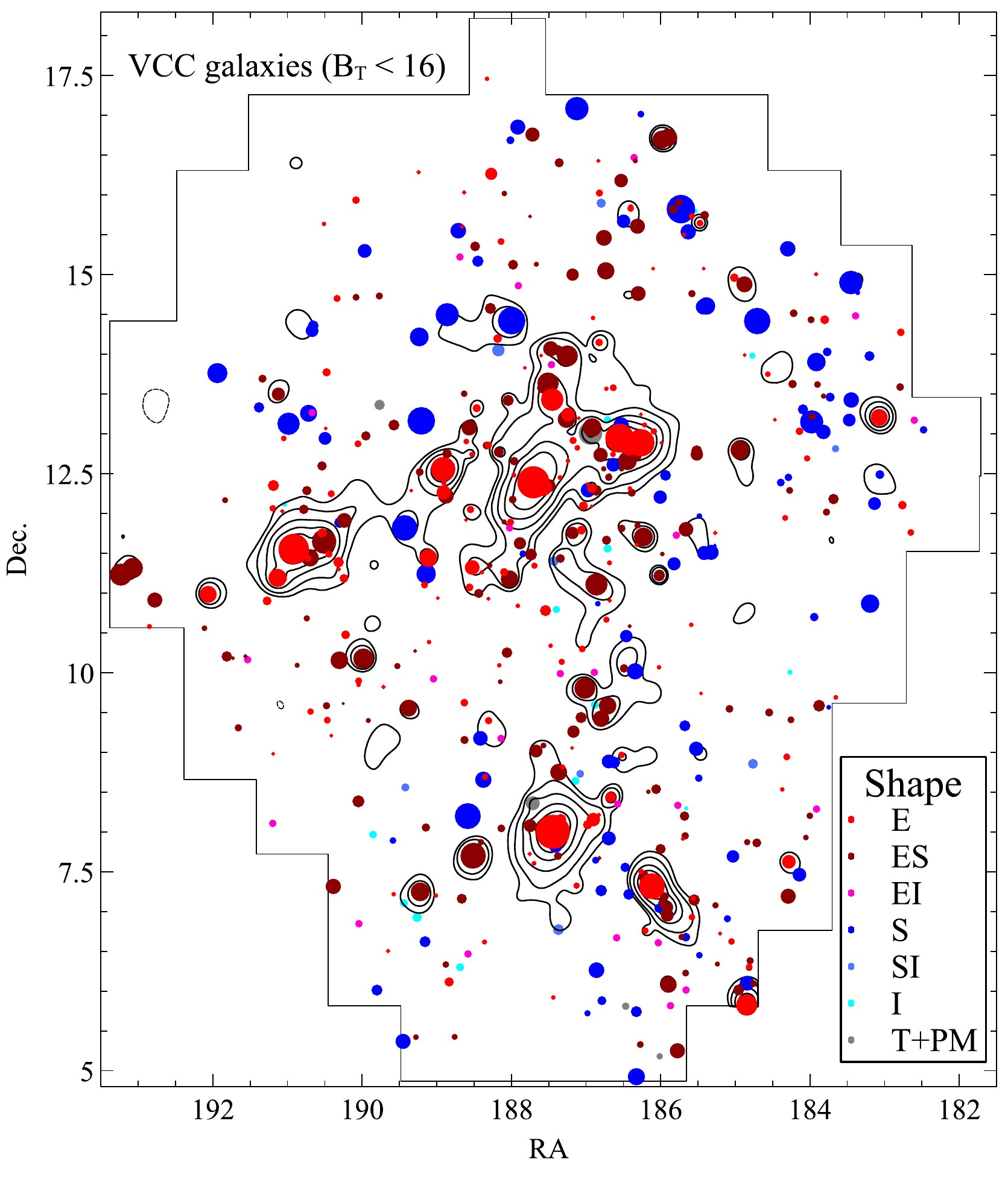}
\caption{Adaptive smoothed GC number density map; the contours are the same as in Figure 9.     The filled circles show the locations of those VCC galaxies within the NGVS with $B_T < 16$; the circle sizes are inversely weighted to $B_T$.    The galaxies are color-coded based on a new classification system using the shapes of the galaxies in deep NGVS images; see text for more details.    E=elliptical ES=elliptical-spiral (transition) EI=elliptical-irregular (transition) S=spiral SI=spiral-irregular (transition) I=irregular T+PM=tidal and/or post merger
\label{fig10}} 
\end{figure*}

\subsubsection{GCs in Virgo Galaxies}      

The observed distribution of Virgo cluster GCs shown
in Figures 9 and 10 largely mirrors that of the distribution of the
luminous early-type (E and ES) galaxies, although the spiral galaxy
M88 (above center) is a strong counter-example.  This is merely the
consequence of early-type galaxies tending to have more GCs per unit
luminosity (the specific frequency $S_N$) than do most spirals
\citep[\eg][]{har91}.

The bulk of the GCs are located in the Virgo A region surrounding M87,
the dynamical center of the Virgo cluster.  Not only are many of the
GCs around M87 distributed along the galaxy major axis
\citep[\eg][]{mhh94,cote01,strad11,forte12}, but the GC distribution
of the entire region appears asymmetric, where more material
is seen to extend to larger distances beyond M87's NW major axis
versus the SE major axis.  While some of the irregular distribution is
due to GCs in the other bright galaxies in the region, we also see
evidence of additional GC populations that will be discussed in
Section 3.3.2.

The Virgo A region around M87 shows many spatial substructures in the diffuse
light \citep{hos05}; our GC maps indicate this is also true for the
GCs.  Such substructure is not only due to the presence of the many
luminous galaxies within Virgo A, but also smaller, less luminous
features that are the result of a continuing and complex evolutionary
history \citep[\eg][]{hos05,jan10,rom12,zhu14}.  In Figure 11 we
superpose our GC contour map over the ultra-deep $V$-band images of Virgo from
\citet{hos05} and \citet{jan10}, which show the distribution of diffuse light as
faint as $\mu_V \sim 28$ mag arcsec$^{-2}$.   The GC contours and the diffuse light surrounding M87 and M49 are very similar, down to levels $\mu_V\sim 27$mag arcsec$^{-2}$.    Such comparisons are difficult at fainter levels due the larger uncertainties in both the GC number densities and the diffuse light.   The smoothing of our maps does not
allow for detection of the many {\it fine} structures and stellar streams
visible in Figure 11 \citep[and described in][]{hos05,jan10}.  The
total luminosities of such streams in both the M87 and M49 regions are
very low (typically$\sim 10^8 L_{\odot}$) and would at most contain a
handful of GCs, although directed spectroscopic studies might prove
interesting.

Also present in the deep $V$-band images are features due not to Virgo Cluster ICL, but to galactic cirrus \citep[see][for more details]{hos05,dav12}.   Comparison with these features (arrowed in Figure 11) with the GC maps show that none are related to enhancements in the GC maps.    Indeed, the magnitudes in our point source catalog have already been extinction corrected according to the \citet{sch98} dust maps based on IRAS 100$\mu m$ maps which also show these brighter cirrus features.

\begin{figure} 
\epsscale{1.20} 
\plotone{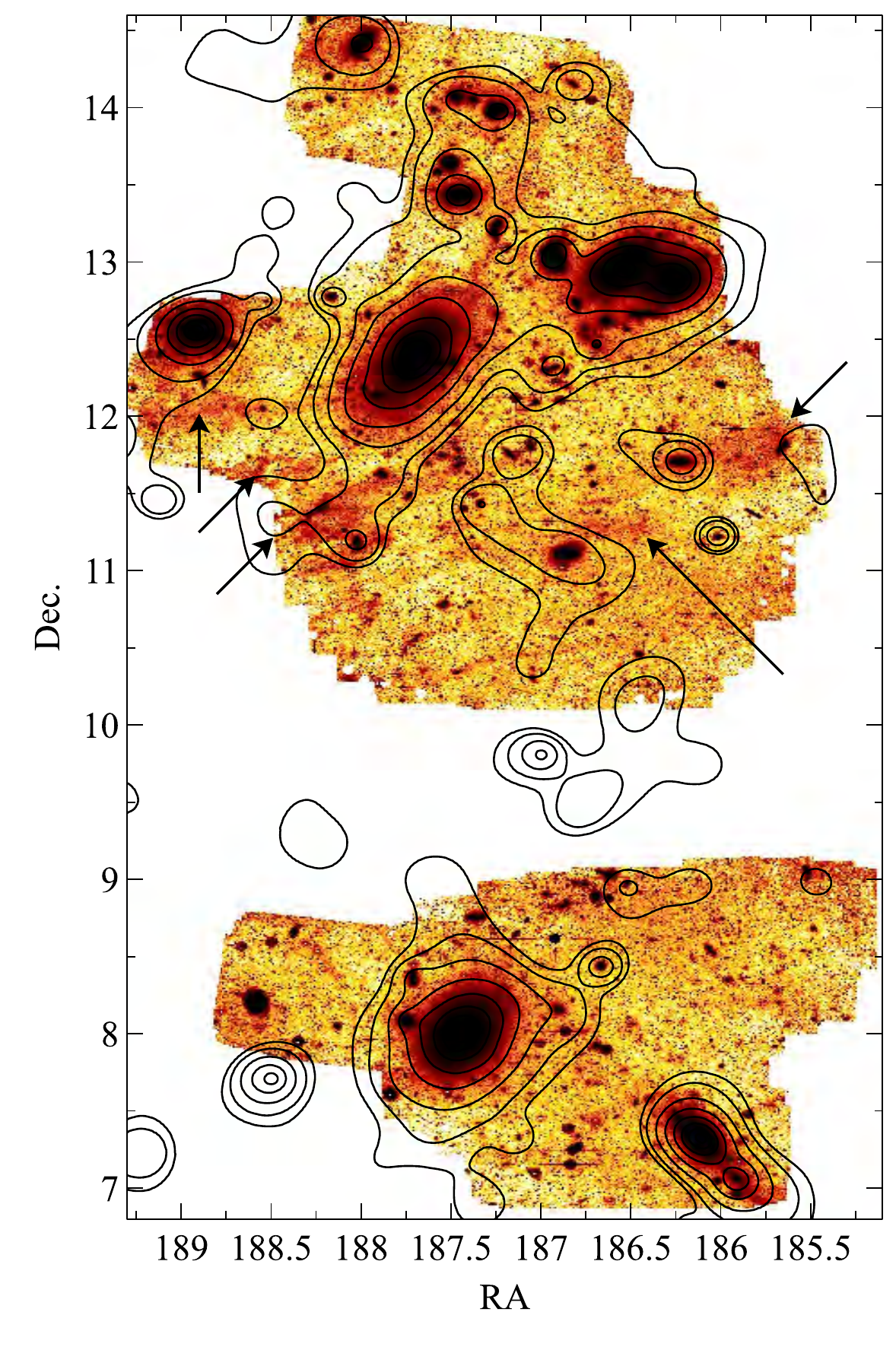}
\caption{The adaptive-smoothed GC map (3, 5, 10, 20, 40, 80$\sigma$ contours) plotted on the deep $V$-band mosaic image from \citet{hos05} and Mihos et al. (2015, in preparation), where the lowest level grayscale represents $\mu_V\sim 28.0$ mag arcsec$^{-1}$.   Note that many of the LSB features below M87 are galactic cirrus (marked by arrows), and not Virgo Cluster ICL\label{fig11}} 
\end{figure} 

\subsubsection{Isolated GCs?}

While most of the GCs shown in Figures 9 and 10 are coincident with known Virgo galaxies, there are a number of regions with no known cluster galaxies that lie at greater than $3\sigma$ above the 
background in the summed bGC$+$rGC map that could represent regions of Virgo with a 'pure' population of IGCs.   These features are labeled I1-I5 in Figure 9.   One such feature (I5) is the GC system of NGC 4651, which is not within the VCC survey area.   The remaining 4 features, however, lie between $3\sigma$ and $4\sigma$ above the background.   As our background maps are based solely on the spatial distributions of two rather different populations -- foreground Milky Way halo stars and background galaxies -- variations in the number densities of either of these populations (\eg background galaxy clusters) could lead to a {\it small} number of spurious $3\sigma$ detections.    We do find one region in the total GC map in Figure 9 that lies $>3\sigma$ {\it below} the background, so it is possible that many (if not all) of these four isolated features are spurious.    Analyses in future papers where the background contamination will be reduced through the additional use of $u^*$ and $K$-band imaging \citep[\eg][]{ngvsir} and/or spectroscopic followup will be required to provide any further insight on these features.

\subsection{The Extended GC Populations of Virgo A (M87) and Virgo B (M49)}

Due to the smoothing applied to create the GC surface density maps, a
detailed look of the spatial distribution of globular clusters around
individual Virgo cluster galaxies is beyond the scope of this paper,
and are left for future papers in the series where either multiple
colors are used to reduce the background noise and/or spectroscopic
confirmation of cluster membership of individual GCs is done.
However, it is illustrative to use our maps to look at the
extent of the GC systems around the two most massive subclusters in
Virgo, shown in Figure 12.    The sheer complexity of the distribution of GCs in
the Virgo A region is presumably a combination of GCs from M87, 
surrounding galaxies (including M84 and M86 to the NW), 
and any possible extended IGC population.  The Virgo B region is also surrounded by the GC systems of many (smaller) galaxies.  We can characterize the outer regions of the GC system of each galaxy through fitting ellipses
to the smoothed maps.

\begin{figure} 
\epsscale{1.10} 
\plotone{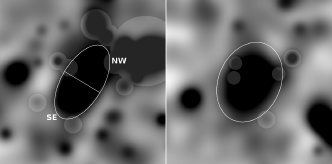}
\caption{Grayscale images of the central  $3\degr \times  3\degr$ region of the adaptive smoothed GC number density maps of M87 (left) and M49 (right).  The masked regions for the ellipse fitting are shown, as are ellipses with 
$r_{maj}=45\arcmin$ (with $\epsilon=0.5$ for M87, and $\epsilon=0.25$ for M49) that represent the galactocentric distance at which the rGC population is consistent with zero; see text for details. \label{fig12}} 
\end{figure} 

\subsubsection{Elliptical Profiles}

For the ellipse fitting, we use the maps with the variable FWHM as
they (a) allow for small smoothing values in regions with high \gc
(\eg the cores of M49 and M87), and (b) have a relatively low noise
$\sigma_{GC}$ to allow us to define the presence (if any) of any extended GC
component around these galaxies.  The use of a spatially variable
FWHM, however, comes with caveats of its own, as the smoothing in the
outer regions of M87 and M49 can be FWHM$\sim30-40\arcmin$, and can
have an adverse affect on some of the measured parameters.  
Where possible we make use of the constant FWHM maps (see Figure 8) to assess our results.

By fitting ellipses to both GC systems, we are making the
assumption that any GCs within the ellipses are (nominally)
associated with each galaxy, and any objects that deviate strongly
from such a regular system would be considered non-members.
Without spectroscopic confirmation of individual GC candidates,
however, such a simplistic assumption is necessary.  Due
to spatial irregularities in the low density outer regions of each
subcluster, we place strong constraints during the ellipse fitting
process.   We used circular masks to remove as many GCs from neighboring
galaxies as possible; the masked regions are shown in Figure 12.
   
We used the ELLIPSE task in IRAF, where an iterative approach to
modeling the elliptical isopleths is carried out for major axis radii
from $r_{maj}=1$ pixel ($0$\minpoint 2) to 350 pixels ($83'$) for both
galaxies on the rGC, bGC and GC adaptive-smoothed maps from Figure 7.
The fits continued until the number of rejected pixels exceeded 50\%
of each ellipse.  The centers of the ellipses were fixed to the
centers of each galaxy.  We also fixed the position angle of the M87 GC
system at PA$=149\degr\pm 4\degr$, as tests showed the major axis of
M87's GC system to be relatively constant with radius.  This value is consistent
with that of M87's light profile \citep{jan10}.  For M49 we allowed
the ellipticity $\epsilon$ to vary, and fixed the
PA$=156\degr\pm3\degr$; again, this PA range for the GC system mirrors
that of the underlying light \citep[\eg][]{jan10}

The derived ellipticities $\epsilon$ for the isopleths are strongly
affected by smoothing with a circular kernel.  To determine outside
what radii such effects are minimized, we performed ellipse fits to
the FWHM$=10\arcmin$ map, where such effects should be minimal for $r
\gtrsim 3\times$FWHM$=30\arcmin$ \citep[\eg][]{schw79}.  We cannot
provide a reliable ellipticity profile of the GC systems interior to
this.  We found that $\epsilon$ derived from both the
adaptive-smoothed and constant smoothed maps for $30\arcmin < r_{maj}
< 50\arcmin$ to be similar : $\epsilon_{M87,GC}=0.50\pm 0.05$ and
$\epsilon_{M49,GC}=0.25\pm0.05$.  We adopt these values for defining
the GC systems in the outermost regions (see Figure 12).  The M87
value can be compared to previous studies that show
$\epsilon_{M87,GC}$ increases with radius
\citep{mhh94,strad11,forte12}, and our results suggest this trend
continues to larger radii.  The resulting number density profiles \bgc, \rgc and \gc are shown in
Figure 13.  The uncertainties are based solely on the rms errors of
the intensity around each ellipse.  The adopted errors are large as the data points are not completely independent of each other, as each ellipse fit is based on the results of the previous ellipse; see \citet{jed87} and \citet{busko96} for technical details of the ellipse fitting algorithm and associated uncertainties.  However, the size of the error bars gives a reasonable view of the uncertainty at any one point, particularly in the outermost regions where the uncertainties due to the background are significant.  Only data points outside
$r_{maj}=3'$ for M87 and $r_{maj}=5\arcmin$ for M49 (corresponding to
the smallest smoothing FWHM in the core region of each galaxy) are
shown, as all surface densities interior to this are smoothed down.
The GC systems at smaller radii around both galaxies have been
previously studied in greater detail
\citep[][]{mhh94,hhm98,kundu99,lk00,tam06b,wat09,har09b,forte12}.  For
comparison, M87's $V$-band light profile (from the Schmidt images of Mihos \etal, in preparation) has also been included.  Previous studies have shown that the surface density profiles of the red GC systems in elliptical
galaxies closely follows that of the underlying light
\citep[\eg][]{geis96}.  We see this trend continue to the outer
regions of both M87 and M49.

\begin{figure*} 
\epsscale{1.00} 
\plotone{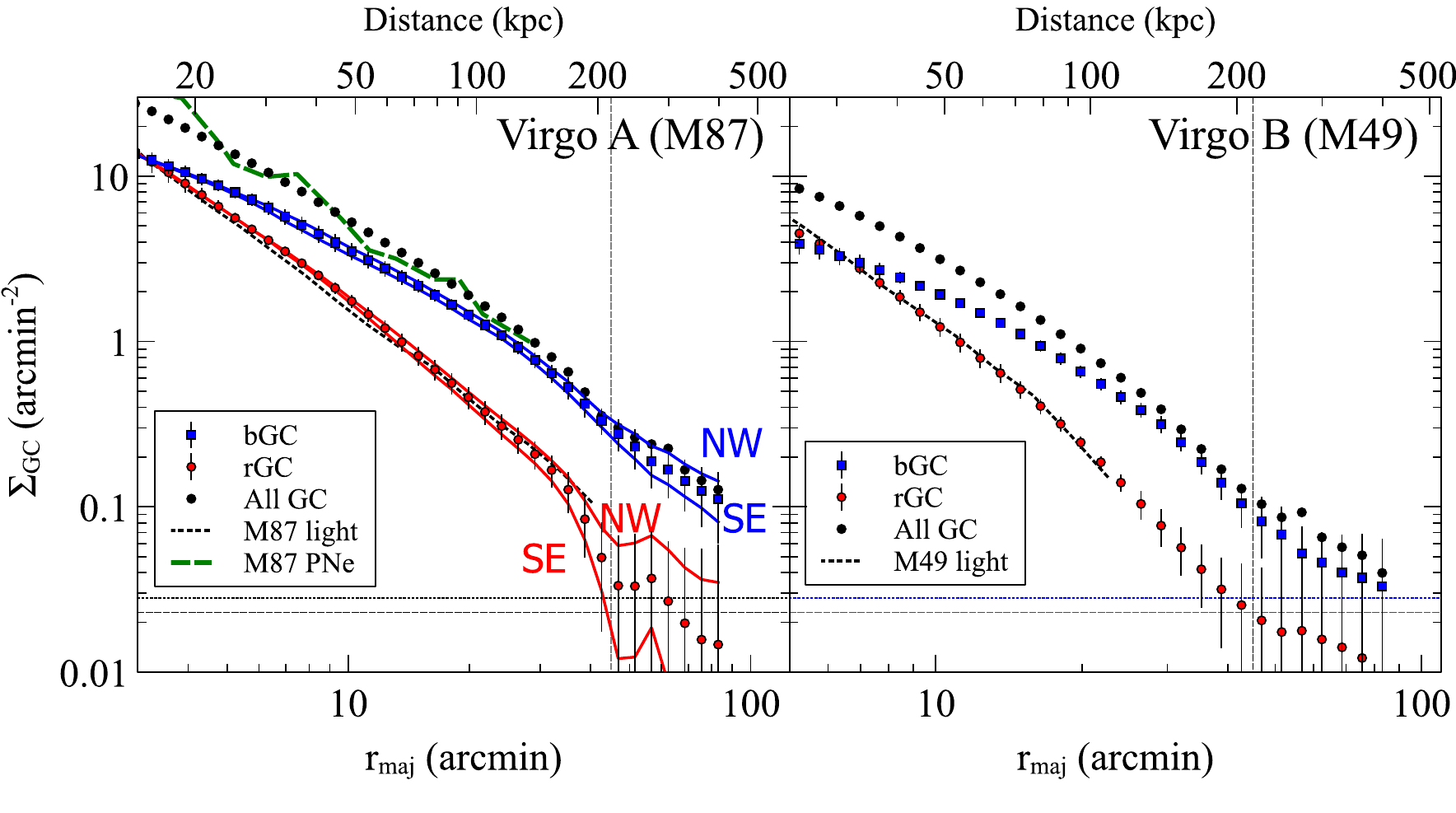}
\caption{Surface density profiles for the globular cluster populations around M87 and M49, based on ellipse fitting to the adaptive-smoothed GC maps.   Shown are the profiles for the red GC population (circles; red in the online version), the blue GC population (squares; blue in the online version) and the summed GC (bGC+rGC) population (black circles).   The black dashed lines denote the $V$-band light profiles from \citet{jan10}, arbitrarily scaled vertically to match the red GC population.   The thick dashed line (green in the online version) denotes the spatial densities $\Sigma_{PN}$ of M87's planetary nebulae (PNe) from \citet{long13}, scaled up by a factor of 5.5 to compare directly with the summed GC profile.   The solid lines denote the profiles resulting from ellipse fitting the NW and SE regions of M87.    The horizontal dashed lines illustrate the $1\sigma$ level based on the uncertainties in the background for the rGC and bGC populations, respectively.  The vertical dashed line denotes the radius $r_{maj}=45'$ outside which the red GC population disappears.   NOTE: Only those parts of the profiles outside $r_{maj}\sim FWHM$ for the smallest smoothing values are shown: $r_{maj}>3\arcmin$ (M87) and $r_{maj}>5\arcmin$ (M49)  \label{fig13}} 
\end{figure*}

\subsubsection{Virgo A (M87)}

The mean M87 GC profiles in Figure 13 show the known
difference in the spatial distribution of GCs around galaxies, where
the blue GCs have a much more extended profile than do the red GCs
\citep[for M87,][]{cote01,tam06b,strad11,forte12}.    Due to the low
surface densities we probe here (and the complete areal coverage provided by the NGVS), we are able to trace out the GC systems to
much larger radii than in most previous studies.     The profiles in Figures 13 show the red GC system
reaches \rgc=0 (within the errors) at $r_{maj}\sim 45\arcmin$ ($\sim
215$ kpc) from M87.  The blue GC population, however, remains
statistically significant all radii, including 2$\sigma$ above the
background at $r_{maj}=83'$.  The mean \bgc profile may change slope
(to \bgc$\sim0.1-0.2$ arcmin$^{-2}$) at $r_{maj}\sim 50-60\arcmin$,
although the uncertainties in the profiles at these radii are large.
This may be an indication of a more extended population of (largely
blue) GCs throughout the Virgo A region; see Section 4.3 for more
details. 

In Figure 14 we compare our {\it total}  (scaled up by a factor of 2.05 to account for 'missed' GCs; see Section 3.1) bGC and rGC number density profiles $\Sigma_{bGC,tot}$ and $\Sigma_{rGC,tot}$ to the red and blue GC profiles from \citet{tam06b}.  The latter profiles were derived from a 2 deg$^2$ rectangular region E from M87 towards M59.  The bGC profiles are in excellent agreement, and the differences in the $\Sigma_{rGC,tot}$
profiles are likely due to the adopted color criteria of the two studies; the $(V-I)=1.1$ cutoff used by Tamura et al. is significantly redder than our $(g'-i')_o=0.8$.  As a result our \rgc profile contains more 'blue'
GCs as defined by Tamura et al., thus making our rGC profile shallower.

\begin{figure} 
\epsscale{1.00} 
\plotone{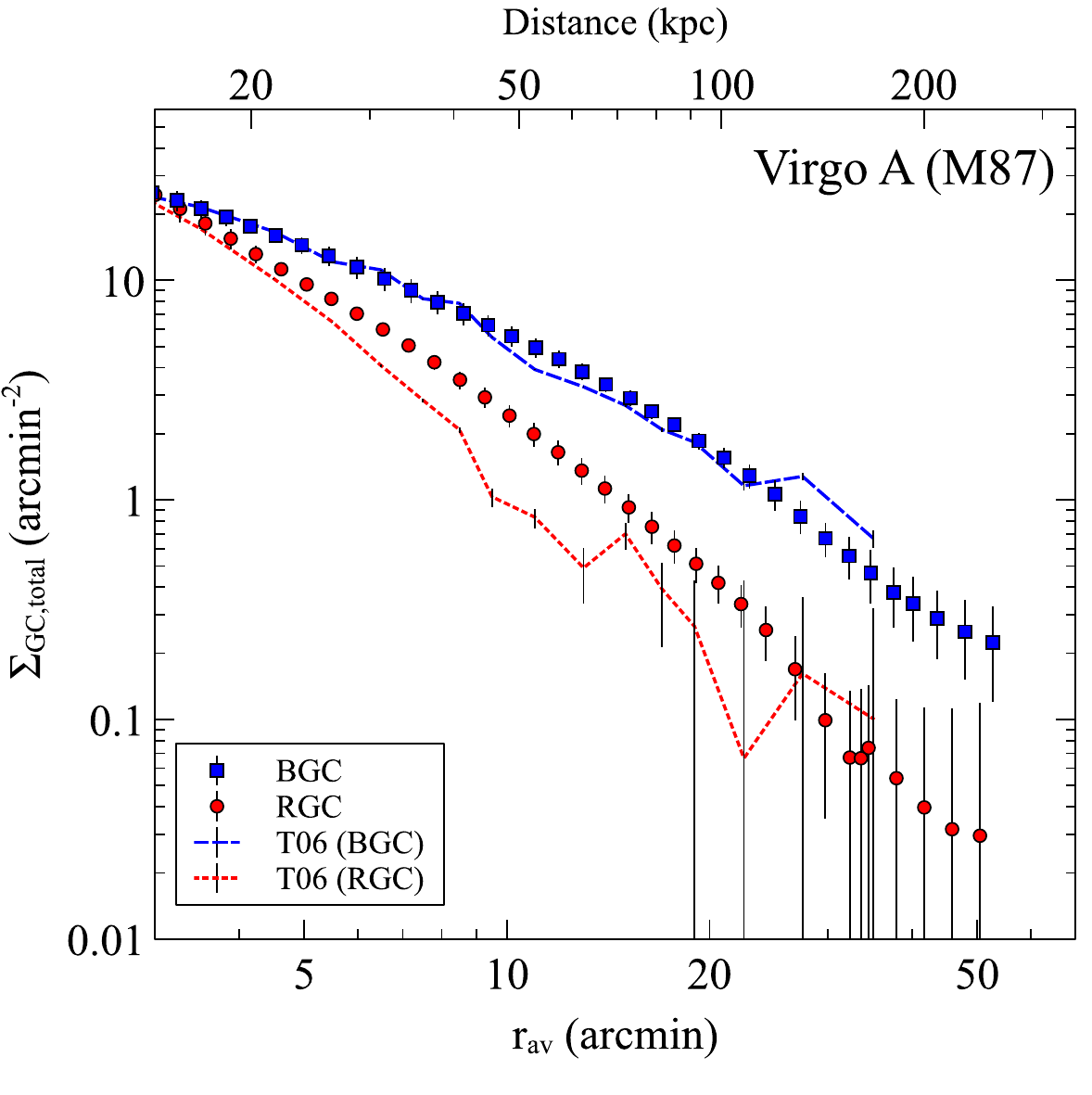}
\caption{Total number density profiles as a function of average radius $r_{av}= r_{maj}\sqrt{1-\epsilon}$ for the blue (squares; blue in the online version) and red (circles; red in the online version) globular cluster populations for Virgo A(M87), derived from ellipse fits to the adaptive-smoothed maps.     Here the plotted number densities $\Sigma_{GC,total}$ have been corrected for those GCs missed in the point-source selection criteria and those objects with \go$>24$.   The dashed lines show the total GC number densities for the blue and red GC populations from \citet{tam06b} for a region stretching east from M87 towards M59.    
\label{fig14}} 
\end{figure} 

To investigate any spatial variations in the M87 GC system, we fit
ellipses to two subsections (NW of M87 along the major axis, and towards the SE) of the M87 GC system in the
adaptive-smoothed maps.  We input the $r_{maj}$, $\epsilon$ and PA
values for the mean profiles (derived above) as input to
ELLIPSE, and derived the surface densities in each half-ellipse.
Here, we allowed the fitting to continue until 50\% of each
half-ellipse was masked out; the results
are plotted as solid lines in Figure 13.  In both the blue and
red GC profiles, we see slightly larger \bgc and \rgc values for the
NW region than observed in the SE region.  To quantify this further,
in Figure 15 we plot the differences in the \bgc and \rgc values
($\Delta \Sigma_{GC} = \Sigma_{GC,NW} - \Sigma_{GC,SE}$) as a function
of radius.  At most radii there is an excess of GCs in
the NW region, although in the inner regions our
adopted uncertainties (based on the rms errors of each of the
half-ellipse fits) are large enough so that any excess is
consistent with zero.  The differences in both the bGC and rGC
profiles are significant (and remarkably constant) at
$r_{maj}>20\arcmin$; for $20\arcmin < r_{maj} < 60\arcmin$ we find
$\Delta \Sigma_{bGC}=0.095\pm 0.015$ and $\Delta \Sigma_{rGC}=0.05\pm
0.01$ (internal uncertainties).    While it is possible that some of
this excess may be due to the presence of overlapping GC systems from
neighboring galaxies (even with the aggressive masking applied before
the fits), this feature is observed at smaller radii where
such interlopers would be even less apparent.  Furthermore, most
galaxies that lie within $r_{maj}=50\arcmin$ from M87 have low luminosities,
and would provide few GCs.

\begin{figure} 
\epsscale{1.00} 
\plotone{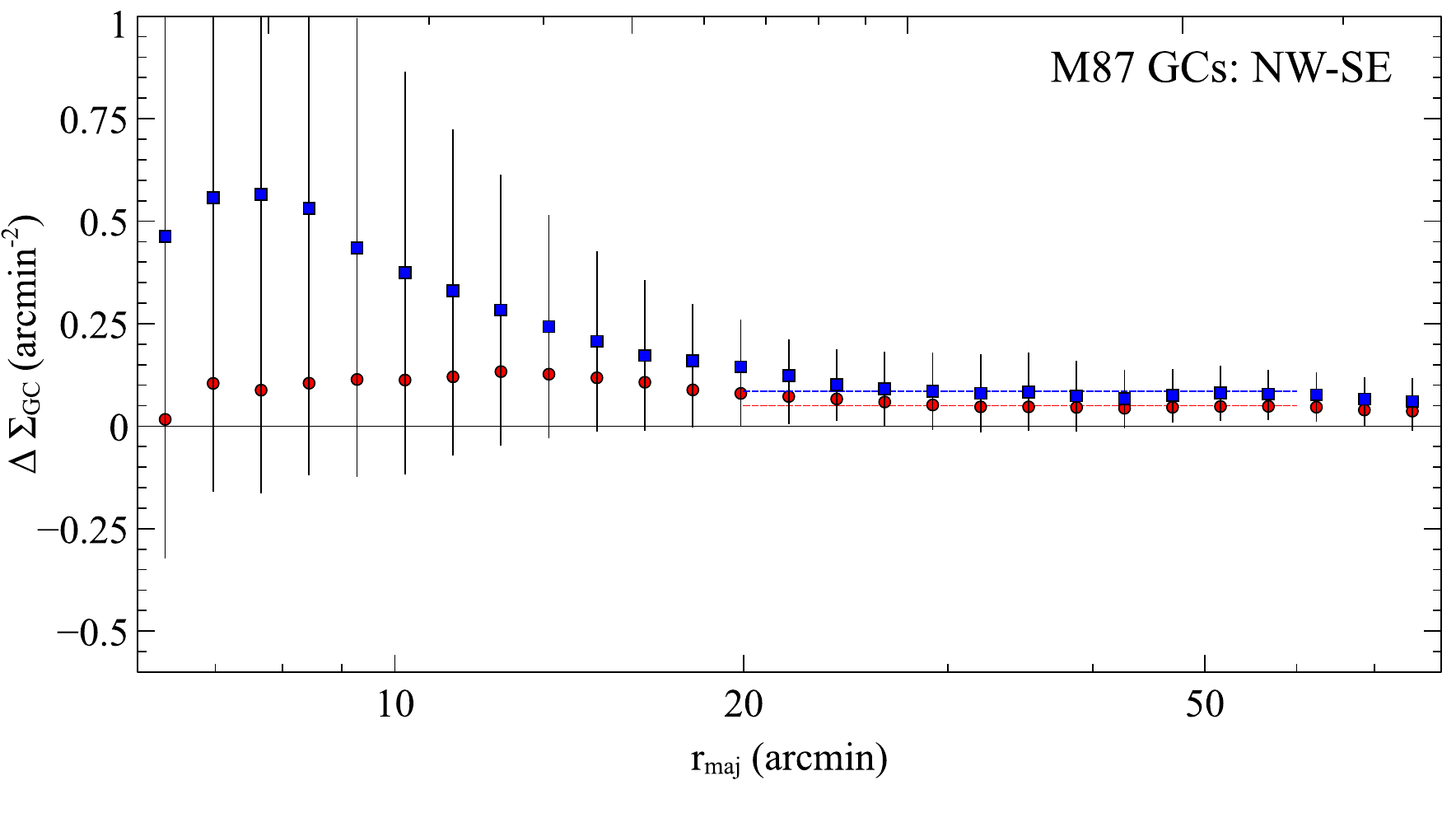}
\caption{Differences in the surface density profiles $\Delta $\bgc (squares; blue in the online version) and $\Delta$\rgc (circles; red in the online version)between the NW and SE regions of Virgo A/M87, as a function of the major axis distance from M87.  The dashed lines show the mean differences $\Delta \Sigma_{bGC}=0.085$ arcmin$^{-2}$ and $\Delta \Sigma_{rGC}=0.05$ arcmin$^{-2}$from points with $20\arcmin < r_{maj} < 60\arcmin$.  \label{fig15}}\end{figure} 

We also note the presence of an enhancement in the GC number density towards the NE, but outside of, the nominal Virgo A GC system described above; this will be discussed further in Section 4.4.

Finally, we examine the deviations from pure elliptical isopleths in
the M87 system.  \citet{hos05} noted that M87's diffuse envelope
appears boxy in its outermost regions, and the total GC map
in Figure 9 shows a similar behavior.  To quantify this, in Figure 16
we plot the $a_4$ term (deviations from a perfect ellipse; the
percentage of the amplitude of the cos$(4\theta)$ term in the Fourier
expansion) as a function of major axis radius derived from the ellipse
fits to both the adaptive smoothed map and the FWHM$=10\arcmin$ map.
In both maps, the $a_4$ profile becomes progressively more negative
(boxy) beyond $r_{maj}\sim 30\arcmin$.  For comparison, we plot the
$a_4$ profile derived from ellipse fits to the deep $V$-band image of
\citet{hos05}, showing the increasing boxiness for $r_{maj}>20\arcmin$
for M87's light.  While the shapes of the M87
diffuse envelope and the surround GC systems are largely similar, we
note that the apparent difference in major axis distance at which
these take place.  However, as noted above, the effect of circular
smoothing on the $a_4$ profile from our maps at smaller $r_{maj}$ is
likely significant.

\begin{figure} 
\epsscale{1.10} 
\plotone{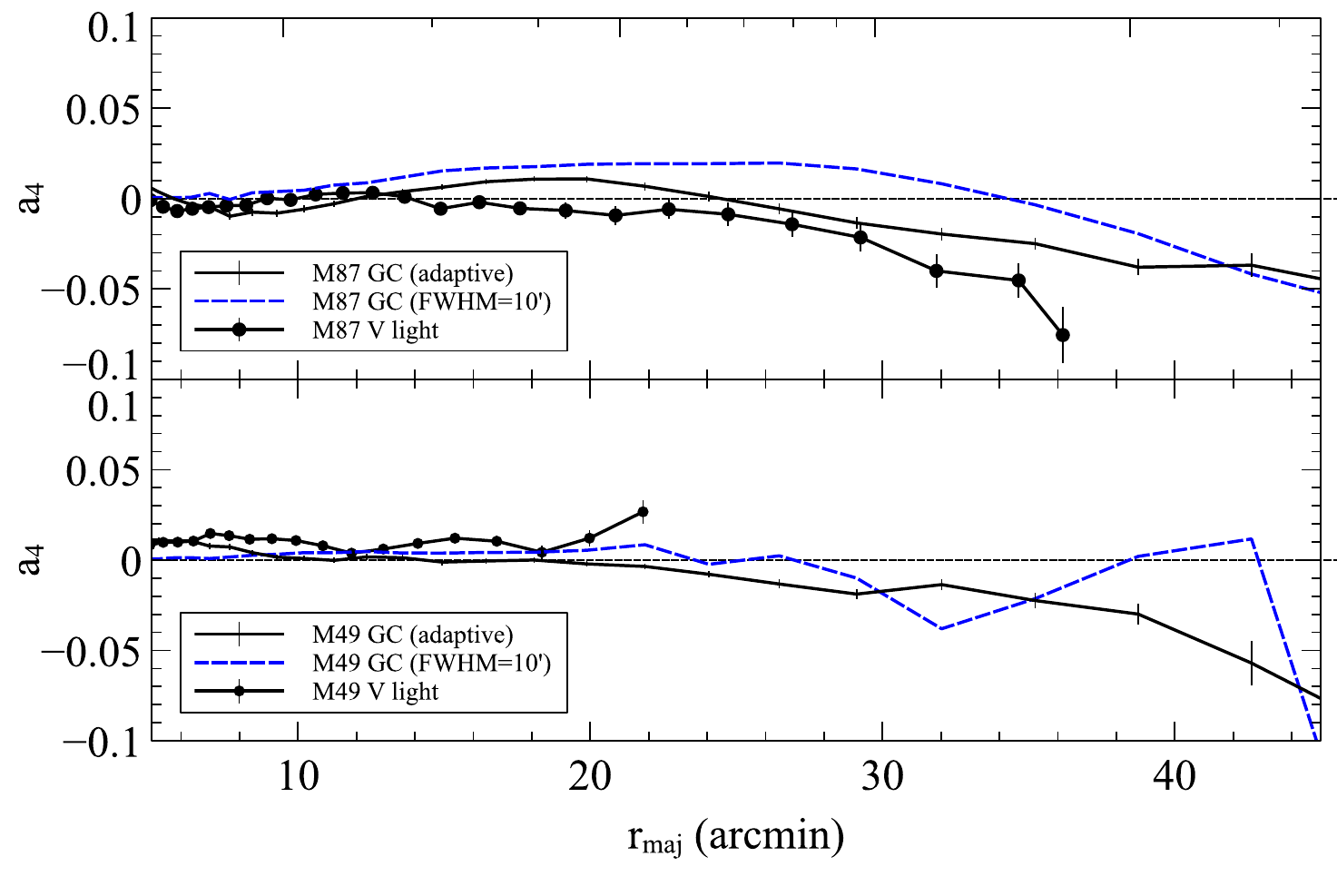}
\caption{The $a_4$ values for the total GC populations of the Virgo A and B systems, plotted as a function of major-axis radius $r_{maj}$.    The solid lines show the $a_4$ profile derived from ellipse fitting to both the adaptive-smoothed map (black line) and the constant FWHM$=10\arcmin$ map (dashed line; blue in the online version).    For comparison, the black circles show the $a_4$ profile for M87's light derived from \citet{jan10}, and for M49's profile from J.C.Mihos \etal, in preparation.  The Virgo A GC systems show increasingly boxy isopleths beyond $30-35'$ (or 140-170 kpc), inside which the $a_4$ values are likely affected by the smoothing. \label{fig16}}
\end{figure} 

\subsubsection{Virgo B (M49)}
Although M49 contains a more prominent system of diffuse 
shells and streams in its outer regions \citep[][]{jan10} than M87,
there is little indication of this in the spatial distribution of GCs in Figure 12, although (as noted earlier) any small GC
populations associated with these features would not appear as significant features in our maps; the six substructures identified by \citet{jan10} have a total
luminosity of $\sim 7\times 10^8 L_{\odot}$, and would optimistically
contribute a total of $\sim 20$ GCs brighter than $g_o =24$.  \citet{cote03} did, however, see hints of an accretion signature 
from M49's GC kinematics.

The \bgc, \rgc and \gc profiles for M49 are shown in Figure 13.  As
observed for the M87 profile, the rGC population matches the $V$-band 
surface brightness profile of the galaxy light quite well, and the rGC profile becomes consistent with \rgc=0 at
$r_{maj}\sim 45\arcmin$ ($\sim 215$ kpc).  The more extended bGC
population continues to much larger radii, and only in the outermost
regions (beyond $r_{maj}\sim 60\arcmin$) does it appear to flatten off
at a level \bgc$\sim 0.05$ arcmin$^{-2}$, or $2\sigma$ above the
background.  If this is due to the existence of a more extended
cluster-wide GC population, the surface density of this population is
lower (roughly half) than that observed around M87.

In Figure 16 we compare the $a_4$ profile of the Virgo B GC system with that derived from the deep $V$-band Schmidt images from J.C. Mihos et al. (in preparation).   While we also see a slight tendency for the Virgo B GC profile to become boxier in the outermost regions, we are only able to compare with the light to $r_{maj}\sim 21'$; however, to this radius, the two profiles {\bf appear} similar.    
            
\subsection{Total number of GCs in the Virgo Cluster}

To derive the total number of GCs throughout the NGVS region, we use
the adaptive-smoothed map from Figure 9.  There is no statistically significant population of GCs in
the outermost regions of the NGVS; the only exception is the
\citep[azimuthally non-isotropic; see][]{ngc4261} GC system of NGC
4261, which partially lies within the NGVS region.  We derive the number of globular clusters $N_{bGC}$ and $N_{rGC}$
through summing the \bgc$\times$(pixel area) and \rgc$\times$(pixel
area) values.  The number of observed blue GCs (\go$< 24$) is
$N_{bGC}=21400\pm 5000$, and that of the red GC population is
$N_{rGC}= 11400\pm 4200$.  The uncertainties are based solely on the
rms errors of \bgc and \rgc from the background fits (the largest
source of error), expanded to cover the 106.5 square degree area of
the NGVS\footnote{This number is larger than the 104 square degrees
quoted earlier due to the inclusion of photometry from areas just
outside the nominal pointings due to dithering.}.  Correcting for the
number of GCs missed by our color, magnitude and size selection
criteria (where $87\%$ of known GCs pass all 3 criteria), these
numbers become $N_{bGC}=24600\pm 5800$ and $N_{rGC}=13100\pm 4800$ for
\go$<24$.     Note that we have made no correction for any faint GCs near the M49 core that may have been missed in the 
SExtractor source lists.  We estimated the total number of missing GCs using the number of velocity-confirmed GCs (13) missing in our source catalog.   Extrapolating to the entire GC luminosity function, we estimate as many as $\sim 100$ GCs (total) are unaccounted for in our estimate.   As this is less than 2\% of the total population, we have ignored their contribution.   

To calculate the {\it total} GC population within our survey area, we
need to correct for those GCs with \go$>24$.  The peak of the GC
luminosity function lies at $g_{o,ACS}=23.9\pm0.2$, where we have
assumed $M_g =-7.2\pm 0.2$ \citep{jor07} and $(m-M)_o = 31.1$ from
\citet{mei07}.    From the relations in \citet{gwyn08}, $g'_{o}
\sim g_{ACS} - 0.1$ for $(g-i)$ colors consistent with GCs.    Thus the
expected peak of the Virgo GCLF is \go$=23.8\pm 0.2$.  Adopting a
symmetric Gaussian of $\sigma_{GCLF}=1.4$ magnitudes\footnote {While $\sigma_{GCLF}$ scales with galaxy luminosity \citep{jor07}, this will not have a significant effect on our results due to our magnitude cutoff lying near the GCLF peak.} for the GCLF, we sample
$56\pm6\%$ of the GCLF.    Correcting for this, our predicted total
number of GCs in the Virgo cluster (bGC and rGC) is $N_{GC,tot}= 67300
\pm 14400$.  As a sanity check, $N_{GC}$ values derived from the (fixed kernel) maps in Figure 7 are within 5$\%$; such differences are far
below the uncertainties due to the background contamination.

\subsection{The M49 and M87 GC Systems}

We also use the \gc profiles from Figure 13 to determine $N_{GC,tot}$
for the individual M49 and M87 GC systems (we did not use the raw \bgc
and \rgc values from the maps due to the masked regions).  For each
galaxy we integrated the mean \bgc and \rgc profiles out to a
major-axis radius $r_{cut}$.  While $r_{cut}$ is difficult to define
for the extended bGC population, any uncertainties in $N_{GC,tot}$ due
to varying $r_{cut}$ are much smaller than the uncertainties that
arise from the correction for GCs below \go=24.  The results are
listed in Table 1.

\begin{deluxetable*}{lcccccc}
\tabletypesize{\scriptsize}
\tablecaption{Globular Clusters in M49 and M87\label{tbl-1}}
\tablewidth{0pt}
\tablehead{
\colhead{} & \colhead{$r_{cut}$ (\arcmin)} & \colhead{$r_{cut}$ (kpc)\tablenotemark{a}} & \colhead{$N_{GC,obs}$ \tablenotemark{b}} & \colhead{$N_{GC,total}$\tablenotemark{c}} & \colhead{$M_{GC,\odot}$\tablenotemark{d} } & }
\startdata
M87 bGC & 60\arcmin & 283 & $4730 \pm 110\pm60$ & $9700\pm1070\pm120$ & $2.3\pm0.3 \times 10^9$ &  \\
M87 rGC & 49\arcmin & 206 & $2350 \pm 50\pm40$ & $4820\pm520\pm70$ & $1.2\pm 0.1 \times 10^9$ & \\
M87 GC &  &  &  & $14520\pm1190\pm140$ & $3.5\pm 0.4 \times 10^9$  &  \\
M49 bGC & 59\arcmin & 283 & $2700 \pm 80\pm100$ & $5540\pm620\pm190$ & $1.4\pm 0.2 \times 10^9$ &  \\
M49 rGC & 49\arcmin & 206 & $1720 \pm 60\pm80$ & $3530\pm400\pm160$ & $0.8\pm 0.1 \times 10^9$ &  \\
M49 GC &  &  &  & $9070\pm740\pm250$ & $2.2\pm 0.2 \times 10^9$ &  \\
\enddata
\tablenotetext{a}{assuming a Virgo distance $d=16.5$Mpc}
\tablenotetext{b}{observed number of GCs to \go$=24$}
\tablenotetext{c}{total predicted number of GCs with the random errors (correction for the missing GCs) and systematic errors (uncertainties in the background) given.}
\tablenotetext{d}{total mass of GCs}
\end{deluxetable*}

Our derived values  of $N_{GC,total,M87}=14520\pm1190$(random)$\pm 140$(systematic) and $N_{GC,total,M49}=9070\pm740$(random)$\pm 250$(systematic) are slightly larger than that found in 
most previously published values \citep[\eg][]{hhm98,rz01,tam06b}, although in most cases this is due to the larger extent in which we are able to trace the GC systems\footnote{\citet{forte12} derive a larger $N_{GC}$ value that we do.  We are unsure of the reason for this discrepancy.  While our values may be slightly underestimated due to a small number of GCs being smoothed to larger radii, this
alone cannot explain the difference.}; when we consider the total number of GCs within smaller radii, our results are consistent with previous works.   Our values are consistent with $N_{GC,M87}=14660\pm891$ and $N_{GC,M49}=7813\pm830$ derived by \citet{peng08} via the integration of the best-fitting Sersic profiles to their GC profiles to larger radii.
   
\section{Discussion}

\subsection{The Virgo Cluster-wide Distribution of Globular Clusters}

Our results indicate that the Virgo cluster has a large GC population,
spread out over a wide area of the NGVS region.  The GCs are distributed, unsurprisingly, much like that of the
early-type galaxies in Virgo.   The
lowest-resolution (FWHM$=30\arcmin$) map in Figure 7 shows the extent
of the (both red and blue) GCs throughout Virgo, down to a $3\sigma$
limit \gc$=0.084\ $arcmin$^{-2}$ (or a total GC number density 
$\Sigma_{GC,tot}=0.17\ $arcmin$^{-2}$), where GCs are seen to extend to
large distances from the large galaxies.    We see GCs $\sim
1\fdg75 - 2\fdg5$ (500-700 kpc) from M87; for the M49 region we see
GCs extend out to at least $=1\fdg25$ ($\sim 350$ kpc).  The populous
GC systems of M49 and M87 together comprise $35\pm 7\%$ of the total
Virgo GC population, even though those galaxies contain only $\sim 8\%$ of the total
$B$ luminosity of the Virgo cluster.   The majority of these GCs are distributed (albeit irregularly) around M87, the center of the most
massive subcluster in Virgo.   As a result of this, the bulk of the GCs
also appear to follow the principal axis of the cluster \citep{wb00}
from the M60 subcluster, westward through M87 to beyond M84/86.  

The distribution of Virgo's GC population is not only similar to that of
the diffuse light present in the cluster, but also closely matches
that of the hot X-ray intracluster gas \citep[also noted by][]{lee10}.  In Figure 17 we compare our
FWHM$=30'$ map contours with the X-ray image from \citet{boh94}, derived from
the ROSAT All-Sky Survey.  The dashed X-ray contour shows the close correspondence between the X-rays and the GCs down to the $3\sigma$ level of the latter.  Due to the limits in the GC number densities due to our background, we cannot effectively compare the two distributions at larger distances where there is still significant X-ray emission.   

\begin{figure} 
\epsscale{1.10} 
\plotone{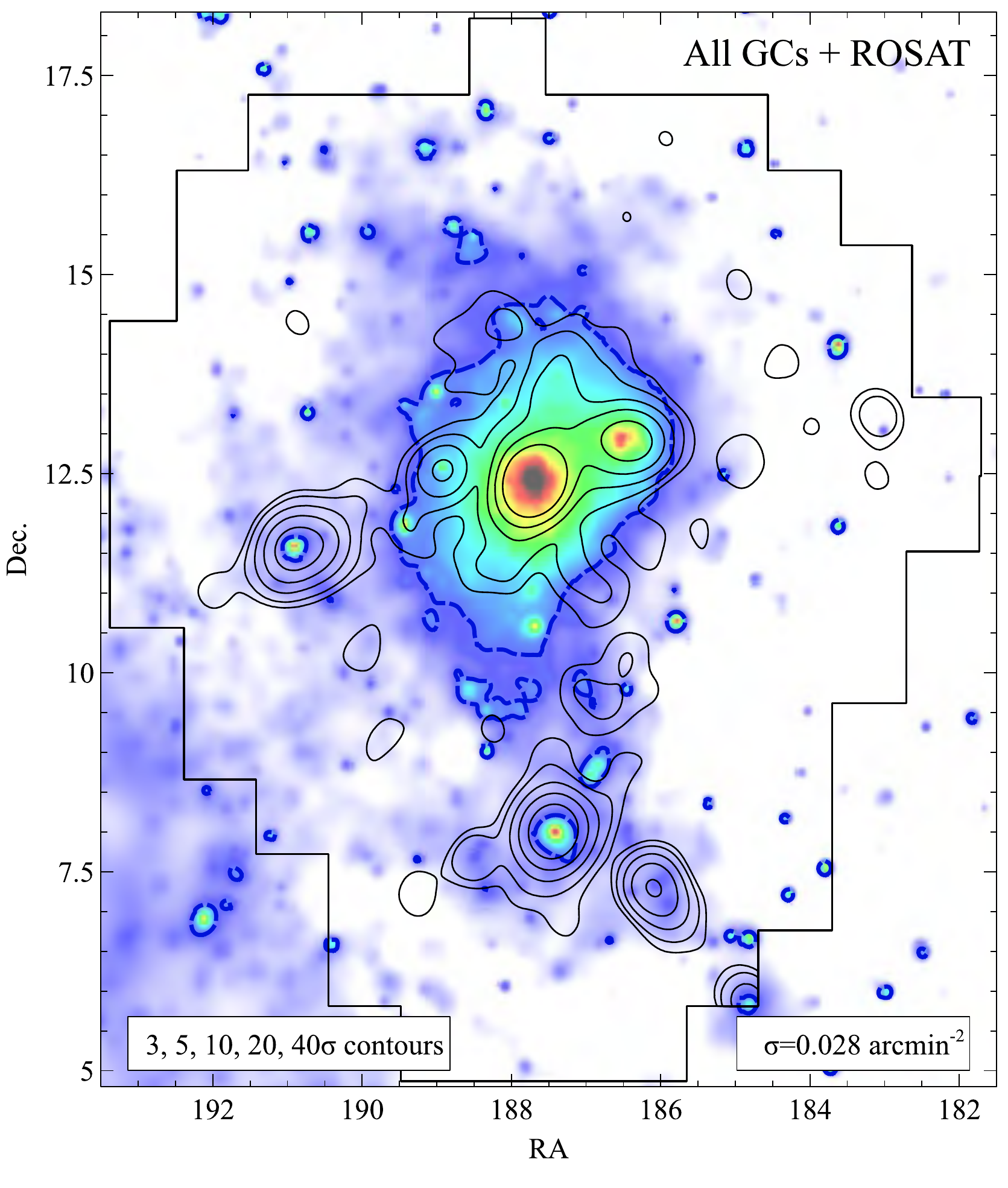}
\caption{The FWHM=30' smoothed map contours (3, 5, 10, 20, 40$\sigma$ above the background) plotted over the ROSAT All-Sky Survey image of Virgo in the $0.5-2.0$keV range \citep{boh94}.  The dashed line denotes the X-ray level at a value $\sim 11\sigma$ above the background, based on inspection of their Figure 1. \label{fig17}} 
\end{figure} 

We not only confirm that the blue GCs in Virgo have a more more
extended distribution than the red GCs, but that the red GC
populations around both M49 and M87 largely disappear (to the limits of our
data) at galactocentric distances $r > 215$ kpc.  {\it Beyond this, the majority
of the GC candidates are (metal-poor) bGCs}.   That we have (largely) metal-poor GCs lying several hundreds of kpc from the center of Virgo does suggest some or most of them are part of the intracluster component of Virgo. We discuss the evidence for such an extended population {\bf below}.   This would indicate that the bulk of the GCs at such large distances from the primary galaxies had originated either in dwarf galaxies \citep[which have preferentially blue GCs,][; see also Schuberth \etal 2010]{peng08}, or from the outer regions of larger galaxies, where fewer red GCs are located.     It is important to note a {\it small} population of rGCs could certainly
exist at larger radii; an extended GC population composed of $\sim 10\pm 10\%$ red GCs \citep[][found $\sim 20\%$ red GCs in the Coma cluster core]{peng11} could still be present within the
uncertainties in our maps.

\subsection{Comparison with Lee et al. (2010)}

\citet{lee10} were the first to study the cluster-wide distribution of Virgo's GC 
population through the use of SDSS photometry and using an adaptive smoothing method similar to that applied here.  They found that GCs
are distributed over a large region of Virgo, with the densest GC
regions (with both blue and red GCs) around the dominant galaxies in
many Virgo subclusters, including M87, M84/86, M60, and M49.  The red
GCs only appear to exist around these major subgroups.  The blue GCs are
much more extended, and fill the region in Virgo along its
principal axis.

The NGVS photometry we use probes much fainter limits to that possible
in that study ($\sim 13\%$ of the GCLF compared to $56\pm6\%$ in our
study), allowing us to detect more than 4 times as many GCs.  More importantly, that we use more GCs to define
our maps not only allows enables us to probe scales of a few
arcminutes, but also facilitates a better characterization of the background 
noise level.

A comparison between Figures 7 and 2(a) from Lee \etal shows generally
good agreement, down to similar number densities.  Their lowest
contour corresponds to (after correction for different photometric
depths) \gc$\sim0.055$ arcmin$^{-2}$, or the $2\sigma$ level in the
FWHM$=30\arcmin$ map, where we find a similar distribution of GCs
around M84/M86/M87/M89, M49, and M60.  However, there are some notable differences between the maps.  We are
able to trace out the GC systems for all galaxies in 'Markarian's
Chain' \citep{mark61} that extend north of M87 out to NGC~4477.  For this
reason (and the detection of M88's GC system) we find GCs extending
over 2 degrees north of M87; this is a feature seen in the X-ray map in Figure 17, but
not in L10.  Similarly, a large, diffuse GC region west of M87 in the
L10 map but not in the X-ray maps is not seen in our maps at all,
although the prominent GC system of NGC~4168 -- not a Virgo cluster
member -- does appear in this area.  

\subsection{Cluster $S_{N}$ and GC Formation Efficiency}

Our value of $N_{GC,tot}$ is the first true measurement of the GC population
over an entire galaxy cluster.  Thus we are able to investigate the {\it
cluster} specific frequency $S_{N,CL}$ as an extension of the usual
$S_N$ (number of GCs per unit luminosity) defined for individual galaxies
\citep{hvdb81}.  To derive the total luminosity $L_V$ of the Virgo Cluster,
we used the $B_T$ magnitudes for all VCC galaxies that are (a) confirmed or
probable members of the Virgo cluster and (b) located within the NGVS
footprint.  We adopted colors $(B-V)=0.95\pm0.10$ for all galaxies
classified E/EI/ES (from Section 3.2 and Figure 10) and $(B-V)=0.70\pm0.20$
for all other galaxies. These values (and their assumed errors) are based on
inspection of colors from the RC3 \citep[\eg][]{rc3} for the brighter early
and late-type galaxies.  From these we derive a total $L_V=1.84\pm0.16 \times
10^{12} L_{\odot}$, for Virgo's galaxies.    To account for any diffuse intracluster light in Virgo that is not included in any VCC magnitudes, we apply an additional (albeit very uncertain) luminosity of $10\pm10\%$ based on studies of Virgo's intracluster RGB stars and PNe \citep[\eg][]{dur02,jjf04b,cr09}.   The total V-band luminosity of the Virgo Cluster is $L_V=2.02\pm0.24 \times 10^{12} L_{\odot}$, or $M_V = -25.95\pm 0.15$ assuming $M_{V,\odot} =+4.81$.  We thus derive $$S_{N,CL}=2.8\pm 0.7$$ for the Virgo
cluster, a value similar to that of elliptical galaxies.  Ignoring the uncertain ICL light fraction would yield a slightly larger $S_{N,CL}=3.1\pm 0.7$.

Many studies \citep{blake97,blake99,sf09,geo10,hhh14} have shown that the ratio of
the GC mass to the total (baryonic+dark matter) mass of galaxies (or GC
formation efficiency $\epsilon_t=M_{GC}/M_{total}$) is similar for galaxies
over a wide range of morphology and luminosity, where $\epsilon_t \sim 4-6
\times 10^{-5}$ \citep{sf09,geo10,hhh14}\footnote{We have changed the results from \citet{sf09} to reflect the median GC mass we have 
adopted}. \citet{mcl99b} found a similar relation relating $M_{GC}$ to the
total {\it baryonic} mass (stars + intracluster gas) in galaxies:
$\epsilon_b=M_{GC}/M_{baryon}\sim0.0026$.  These relationships may reflect
the early formation of the GCs in proportion to total available mass, before
feedback mechanisms become effective in shutting off future star
formation.  \citet{a-m13} noted, however, that such formation efficiencies
are (necessarily) dependent on the radius out to which the various masses
are derived, and found both $\epsilon_b$ and $\epsilon_t$ to decrease with
radius in the massive galaxy cluster Abell 1689.

It is interesting to see if the scaling of GCs with mass for individual
galaxies holds for the Virgo cluster as a whole.  Assuming an average GC
mass $M=2.4\times 10^5 M_{\odot}$ \citep{mcl99b}, the total mass of GCs in
Virgo is $M_{GC} = 1.6\pm0.3 \times 10^{10} M_{\odot}$.  For the baryonic
mass, we combine the intracluster gas mass of the M87, M49, and M86 groups
$M_{gas}=2\times10^{13} M_{\odot}$ \citep{schind99} with the total stellar
mass $M_* = 8.1\times 10^{12} M_{\odot}$, found by assuming $M/L_V=4$ for
old, metal-rich populations \citep{a-m13}.  Thus for a total Virgo baryonic
mass of $2.8\times 10^{13} M_{\odot}$, we find a baryonic GC formation
efficiency $$\epsilon_b=5.7\pm1.1 \times 10^{-4}$$  This value is lower than
$\epsilon_b=2.6\times 10^{-3}$ from \citet{mcl99a} and $\epsilon_b=9.5\times
10^{-4}$ from \citet{a-m13}, although these values are derived at radii of
100 kpc and 400 kpc from the central galaxy, while our results are for the region within Virgo's virial radius ($\sim 1.5$Mpc). For the total cluster mass, we
combine the dark matter halo masses of the Virgo A and B subclusters from
\citet{ngvs}, based on \citet{mcl99a}, and the total mass for the M86
subcluster from \citet{schind99}, to get $M_{Virgo}=5.5\times 10^{14}
M_{\odot}$.  Thus we derive $$\epsilon_t = 2.9\pm0.5\times 10^{-5}$$  where the error is based on the uncertainty in $M_{GC,tot}$ only. This result is
slightly lower than previously derived galactic values of $\epsilon_t=4.2
\times 10^{-5}$ \citep{sf09} and $\epsilon_t=3.9\pm0.9\times 10^{-5}$
\citep{hhh14}, but is well within the observed spread seen in both studies, where many galaxies have $\epsilon_t\sim 1-10\times 10^{-5}$.     Moreover, given the
decline in $\epsilon_t$ with radius \citep{a-m13}, there should be some
limited region within Virgo over which $\epsilon_t$ matches the mean for
individual galaxies, although the global value is lower.   It is important to note that our values of $\epsilon_b$ and $\epsilon_t$ are largely independent of our assumed correction for Virgo's ICL, as Virgo's stellar mass is a small fraction of the baryonic and total masses.

\subsection{Virgo's Extended Stellar Populations}

The extended spatial extent of Virgo's GC population and its striking
similarity to the distribution of the intracluster gas suggests that
some GCs {\it must} be members of the intracluster medium.    Unfortunately, with spatial information alone it is difficult to
distinguish between 'galactic' GCs and IGCs; see \citet{west11} for a
cautionary discussion of IGCs in Abell 1185.  A classic definition of
intergalactic populations are regions where the surface
brightness/number density profile of a galaxy changes slope in the
outer regions, suggesting an additional stellar population.  The
total blue GC number densities in the outermost regions of
Virgo A and B are at the \bgct$\sim 0.2$ arcmin$^{-2}$ and \bgct$\sim
0.1$ arcmin$^{-2}$, respectively.  These values are consistent with the broad limits on any IGC
population by \citet{tam06b}.  That these values are only $2-3\sigma$
above our background level reduces our ability to effectively model
the IGC population as merely an excess of objects above a 1-D profile
of the galactic GC populations (see Figure 13); indeed, we only see
possible evidence for an inflection point at $r_{maj}\sim 50'-60'$
(240-290 kpc) in the \bgc profiles.

Any IGCs superimposed on the large galaxies themselves would only
become obvious through velocity information derived from spectroscopic
follow-up studies -- otherwise they appear as if they were part of the galactic GC system\footnote{This situation is
not unique to IGCs, as there have been many definitions used as to
what constitutes the intracluster component in diffuse light studies
\citep[\eg][]{dolag10,rud11}.}.  As noted in section 3.3.2,
it is possible to find extra components through their spatial
distribution alone.  Modeling (and masking) the GC systems around
individual Virgo galaxies \citep[][for the Coma
cluster]{peng11} is beyond the scope of this work; and such a task will
(necessarily) be left to future studies using multiple colors \citep[\eg][]{ngvsir} and spectroscopy to significantly reduce the 
background contamination.   

With these caveats in mind, we have found a small spatial asymmetry in the GC number densities within the M87 system  (at $r_{maj}<50'$).  We also find possible evidence for an excess in the GC surface densities  NE of M87 (labeled 'NE' in Figure 9) which lies outside the M87 GC system, but in a region with only a few low luminosity dwarf galaxies.      While this latter feature lies $\sim 2.5\sigma$ above the nominal M87 GC system in our maps (5 $\sigma$ above the mean background), it is something to be investigated in future studies.   Both of these spatial features could plausibly be due to the presence of an extended intracluster GC population.   Recent dynamical studies of GCs in the
M87 region have shown additional substructures within 100 kpc from the center of M87 \citep{rom12,zhu14}.   While it is possible that the NW/SE asymmetry we see is related to either of these features, the GC excess we see lies further ($r_{maj}>20\arcmin-50\arcmin$ = 100-250 kpc) from M87 than the kinematic features.   The presence of both spatial and kinematic substructure in the GC populations strengthens the case that the
production of the ICL (including the liberation of the intracluster stellar populations) is an ongoing process that continues to the present time.   

That most of the spatial and kinematic substructures (thus far) are
located towards M87's NW major axis may not be a coincidence.  The
diffuse optical light observed by \citet{hos05,jan10}, the kinematics of
PNe NW of M87 from \citet{doh09} and the presence of GCs extending over 700 kpc NW of M87 are all suggestive of an intracluster population that is not centered on M87.    Although this has
been attributed to the presence of a significant diffuse light
component around M86 (and not M87) by \citet{doh09}, the wealth of
(tidal) substructure around M87 would suggest a larger extended GC population in the Virgo A region.  This can only be
tested by much larger spectroscopic samples in this region of the
Virgo cluster.

A useful comparison can be made between the number density profiles of the GCs in the Virgo A region with that of the PNe.   A spectroscopic study of a small number of PNe in the Virgo core by \citet{doh09} suggests that PNe outside $r_{maj}\sim 210$ kpc\footnote{We have corrected the \citet{doh09} circular-averaged radius $R=149$ kpc to a major-axis radius assuming $\epsilon=0.43$ and using $d=16.5$ Mpc.} from M87 exhibit kinematics expected for an intracluster population, rather than
that of objects bound to M87's halo.    In Figure 13 we overlay the surface density profile of a photometrically selected sample of PNe from \citet{long13} on our mean GC (bGC+rGC) profile.     Although the uncertainties in the PNe profile are not explicitly given (the PNe profile is very sensitive to the background corrections applied), there is excellent agreement in the profiles between the two populations, even though the GCs and the PNe may not necessarily track the same stellar populations.     While \citet{long13} attribute their excess of PNe (compared to that of the galaxy light) to the combination of an M87 halo and
an ICL stellar population (the latter of which has a more metal-poor composition) it is possible that some of their ICL component may
be related to the blue GC component of M87's halo. 

\subsection{Virgo's Dynamical History}

The wealth of both spatial and kinematic substructure in the galaxy population of the Virgo Cluster \citep[\eg][]{bing87,bing93,mei07,paudel13} is evidence the cluster is not in a fully relaxed state.   Furthermore, the observations of Virgo's extended stellar populations (GCs, PNe, diffuse light) discussed above allow a qualitative discussion on the dynamical state(s) of the Virgo Cluster as a whole.   Virgo is the first cluster where the spatial distribution of the entire GC system can be compared to both the cluster-wide intracluster gas and the galaxy population \citep[\eg][]{lee10}.    The 2D distribution of VirgoÕs GCs over a larger scale largely follows the irregular distribution of the galaxies.   This result, combined with the spatial and kinematic substructures observed in the stellar populations both within and outside Virgo galaxies \citep[\eg][]{hos05,doh09,jan10,lee10,rom12,zhu14}, is a clear indication of the recent (and ongoing) ICL/IGC production.    \citet{rud09} suggest that build-up of the ICL is through continual, though episodic processes that have occurred over a large fraction of a Hubble time.  Our results are consistent with this picture.  There are (related) suggestions that a large fraction of the current ICL was pre-processed by stripping from galaxies in the group environment before their eventual infall into the cluster environment \citep{hos04,rud06}.   While we cannot directly test this idea here (as we are not studying the group environments at the lowest level), that the smaller subclusters in Virgo (\eg M49, M60) show less evidence for any IGC population (based on the extent of the bGC populations around M49 and M60 in Figure 7) suggests that ICL production around these (less massive) subclusters is perhaps still in its earliest stages.     Forthcoming $K-$band imaging of Virgo \citep[\eg the methods from][]{ngvsir} promises to reduce the background contamination further and allow us to trace the GC population to lower levels than possible here, and would facilitate comparisons with the predictions of cosmological models \citep[\eg][]{moore06}.     In contrast, the similarity in the spatial distributions of the intracluster gas and GCs around M87 suggests that the two components have similar dynamical histories.  This would indicate that some fraction of the IGC population in VirgoÕs largest subcluster was liberated early on, or at least at early enough times to allow for partial virial equilibrium.      It will be interesting to compare the large-scale GC and X-ray distributions in other clusters in order to determine if a similar correspondence holds in systems of different masses and/or dynamical states, and if we can directly trace the build up of the ICL component through the GCs.

\section{Summary}

We have used $g'i'$ photometry for point sources from the NGVS to extract the population of globular
clusters throughout the survey area.  We show the following:

-- Number density maps of Virgo's GCs show a complex 2-D
structure surrounding the large early-type galaxies in the cluster.
Many of the GCs are located near the core of the Virgo cluster, the
massive subcluster Virgo A.  The Virgo cluster contains a total
population of $67300\pm 14400$ GCs, with a significant fraction ($\sim
35\%$) of these lying within the GC systems of M87 and M49 alone.    
The resulting cluster-wide specific frequency $S_{N,CL}$ is $2.8\pm0.7$, and we derive a GC -to-baryonic mass ratio $\epsilon_b=5.7\pm1.1 \times 10^{-4}$, and a total GC mass ratio $\epsilon_t = 2.9\pm0.5\times 10^{-5}$ within Virgo's virial radius. 

-- The GC systems of M87 and M49 are very similar in shape and extent to the
diffuse light that is known to surround both galaxies.  The M87 GC
system, in particular, shows the same 'boxy' isopleths as seen in the
diffuse light in the outer regions of M87, indicative of a common
origin.  

-- The GC systems in both the Virgo A (M87) and Virgo B (M49)
subclusters extend to large radii.  The red (metal-rich) GCs are
found to follow the galaxy light and are seen to extend as far as
$\sim 215$kpc from the central galaxies; beyond this distance the red
GC population is consistent with zero.  The blue (metal-poor) GC
population has a shallow distribution and extends to much
larger distances, with significant populations of GCs still present
(with total number densities \gc $\sim0.1-0.2$arcmin$^{-2}$) 400 kpc from the central galaxies.  If some (or all) of
the GCs at these radii are 'intracluster' GCs, this suggests such a
population is largely comprised of metal-poor objects, and is 
suggestive of an origin from either dwarf galaxies, or preferentially
removed from the outer regions of larger galaxies.  The gravitational
processes that have liberated these GCs from their original galaxies
continue to the present day.

-- We have found a small asymmetry in the M87 GC system, where an
observed excess of GCs at $\Sigma_{GC,tot}\sim 0.2$arcmin$^{-2}$ is observed
towards the NW of M87.  Whether this feature is caused by a population
of GCs that is offset from, but superposed on, the M87 GC system, or
is partially related to kinematic substructures detected within the
M87 GC system itself \citep[][]{rom12,zhu14} cannot be clearly defined
from our data.    The increasing complexity of the M87 region (both spatially and kinematically) provides strong evidence of the recent and continuing 
processes of galactic accretion throughout the Virgo core.

-- The GCs surrounding M87 have a similar spatial distribution to that of the (relaxed) intracluster gas in the region, suggesting that the build up of the extended GC population began at an epoch 
early enough for some of the GC population to be in a similarly relaxed state.  This
is consistent with the slow, gradual build up of the outer regions of
BCGs, including the ICL.

\acknowledgments

The authors would lie to thank Hans B{\"o}hringer for the use of the ROSAT X-ray data, and the anonymous referee who made numerous useful suggestions to improve this paper.   PRD gratefully acknowledges support from NSF grant AST-0908377.  PRD thanks John Feldmeier and Myung Gyoon Lee for useful discussions throughout the writing of this paper, Jay Strader for making his M87 GC catalog available before publication, and the Department of Astronomy at Case Western Reserve University, where parts of this paper were written.  JCM has been supported through NSF grants AST-0607526 and AST-1108964.  CL acknowledges support from the National Natural Science Foundation of China (Grant No. 11203017, 11125313 and 10973028).  PAD acknowledges support from Agence Nationale de la Recherche (ANR10-BLANC-0506-01).  

The NGVS team owes a debt of gratitude to the director and the staff of the Canada-France-Hawaii Telescope, whose dedication, ingenuity, and expertise have helped make the survey a reality.   This work is supported in part by the French Agence Nationale de la Recherche (ANR) Grant Programme Blanc VIRAGE (ANR10-BLANC-0506-01), and by the Canadian Advanced Network for Astronomical Research (CANFAR) which has been made possible by funding from CANARIE under the Network-Enabled Platforms program. This research used the facilities of the Canadian Astronomy Data Centre operated by the National Research Council of Canada with the support of the Canadian Space Agency. The authors further acknowledge use of the NASA/IPAC Extragalactic Database (NED), which is operated by the Jet Propulsion Laboratory, California Institute of Technology, under contract with the National Aeronautics and Space Administration, and the HyperLeda database (http://leda.univ-lyon1.fr).


{\it Facilities:} \facility{CFHT}

\clearpage

\end{document}